# Shear-enhanced Liquid Crystal Spinning of Conjugated Polymer Fibers

Hao Jiang[1#], Chi-yuan Yang[2#], Deyu Tu[2], Zhu Chen[3], Wei Huang[4], Liang-wen Feng[3], Hengda Sun[1], Hongzhi Wang[1*], Simone Fabiano[2*], Meifang Zhu[1], Gang Wang[1*]

[1] State Key Laboratory for Modification of Chemical Fibers and Polymer Materials, College of Materials Science and Engineering, Shanghai Key Laboratory of Lightweight Structural Composites, Ministry of Education Key Laboratory of High-Performance fibers & products, Donghua University, Shanghai 201620, P. R. of China.

[2] Laboratory of Organic Electronics Department of Science and Technology, Linköping University, Norrköping SE-60174, Sweden.

[3] Key Laboratory of Green Chemistry & Technology, Ministry of Education, College of Chemistry, Sichuan University, Chengdu, 610065, China.

[4] School of Automation Engineering, University of Electronic Science and Technology of China (UESTC), Chengdu, P. R. China.

**Corresponding Author**: Gang Wang, E-mail: gwf8707@dhu.edu.cn

Simone Fabiano, E-mail: simone.fabiano@liu.se

Hongzhi Wang, E-mail: wanghz@dhu.edu.cn

**Abstract**

Conjugated polymer fibers can be used to manufacture various soft fibrous optoelectronic devices, significantly advancing wearable devices and smart textiles. Recently, conjugated polymer-based fibrous electronic devices have been widely used in energy conversion, electrochemical sensing, and human-machine interaction. However, the insufficient mechanical properties of conjugated polymer fibers, the difficulty in solution processing semiconductors with rigid main chains, and the challenges in large-scale continuous production have limited their further development in the wearable field. We regulated the π-π stacking interactions in conjugated polymer molecules below their critical liquid crystal concentration by applying fluid shear stress. We implemented secondary orientation, leading to the continuous fabrication of anisotropic semiconductor fibers. This strategy enables conjugated polymers with rigid backbones to synergistically enhance the mechanical and semiconductor properties of fibers through liquid crystal spinning. Furthermore, conjugated polymer fibers, exhibiting excellent electrochemical performance and high mechanical strength (~600 MPa) that essentially meet the requirements for industrialized preparation, maintain stability under extreme temperatures, radiation, and chemical reagents. Lastly, we have demonstrated logic circuits using semiconductor fiber organic electrochemical transistors, showcasing its application potential in the field of wearable fabric-style logic processing. These findings confirm the importance of the liquid crystalline state and solution control in optimizing the performance of conjugated polymer fibers, thus paving the way for developing a new generation of soft fiber semiconductor devices.

**Main**

Owing to their π-π conjugated structures, conjugated polymer exhibit superior carrier transport capabilities, thus offering wide-ranging applications in areas such as flexible electronics, semiconductor photovoltaics, and biosensors[1–4]. Recent developments in conjugated polymer have imparted diverse electronic functionalities to fibers and textiles, making soft fiber electronic devices attractive for human-machine interaction, personalized healthcare, and energy conversion[5–8]. Fibers and textile devices capable of generating, transmitting, modulating, and measuring electronic functionalities represent the more potential and easily accepted wearable electronics form at present. Fibers used in commercial garments, such as cotton and jute, need to be mechanically strong to meet the requirements of mechanized production. However, the typical tensile strengths (cotton>400 MPa; jute>350 MPa) still pose challenges to the electronic fibers. The inherent rigidity of the backbone and melting temperatures near thermal decomposition points make the preparation of conjugated polymer fibers through electrospinning and melt spinning exceedingly difficult[9–11]. For wet spinning, PEDOT:PSS and PANi fibers have already achieved enhanced mechanical and electrochemical properties through post-spinning drawing and innovative solvent exchange methods[12,13].

The liquid crystal state of conjugated polymers is an equilibrium intermediate phase driven by π-π interactions and steric hindrance effects[14]. During the liquid-to-solid transition in fiber spinning, the ordered aggregate structure formed in the solution is inherited by the solid-state fiber[15]. This orientation usually exists only at the microscopic level, while at the macroscopic level, the liquid crystalline materials exhibit isotropy[16]. By applying fluid shear stress for secondary orientation induction in ordered liquid crystal aggregates, the close stacking of macromolecules in fibers is enhanced, reducing internal flaws. This facilitates achieving high orientation and crystallinity, thus synergistically enhancing the carrier transport and mechanical properties of one-dimensional structural fibers. During the liquid crystal formation process, the transition of lyotropic liquid crystal polymers from isotropic dispersions to liquid crystal dispersions is driven by changes in concentration[17]. However, the lower solubility and inappropriate rheological properties of conjugated polymer s make it difficult to match the processing window of the spinning process. This poses a challenge in producing high-quality conjugated semiconductor fibers using conventional liquid crystal spinning techniques[18].

Herein, we employed fluid shear stress to achieve continuous liquid crystal spinning of several typical conjugated polymers, by promoting π-π stacking interactions to prepare semiconductor fibers with high orientation and crystallinity. Combining X-ray diffraction with electrochemical analysis, we quantitatively depicted the influence of shear stress on the π-π stacking, crystallinity, and charge transport characteristics of conjugated semiconductor fibers spun from liquid crystals. We also discovered that the spontaneous orientation of liquid crystal molecules and the secondary orientation enhanced by fluid shear led to a significant axial preferential orientation in conjugated semiconductor fibers, resulting in anisotropic electrochemical properties, with axial carrier mobility and transconductance enhanced by ~400%, respectively, compared to the radial direction. Conjugated semiconductor fibers, demonstrating exceptional electrochemical properties, also possess high mechanical strength (~600 MPa) and stability in extreme temperature, radiation, and chemical conditions, indicating their potential in creating practical and highly adaptable logic fabrics. Lastly, we demonstrated a NAND logic circuit based on semiconductor fiber organic electrochemical transistors (OECTs), proving its tremendous potential in fabric-level logic chip applications. This discovery breaks through the limitations of traditional conjugated polymer s and provides new directions and inspiration for developing high-performance, wearable electronic devices.

**The continuous liquid crystal spinning of semiconductor fiber.**

Poly(benzimidazobenzophenanthroline) (BBL) was chosen as an example of liquid crystal conjugated polymer with intermolecular interactions and a highly rigid backbone structure. In this study, the spinning fluid was pressure-driven through a microfluidic channel, and shear forces enhanced the orientation and phase transformation of the liquid crystal conjugated polymer. Due to its highly oriented molecular chains, liquid crystal phase structure, and fast proton exchange process, BBL can be spun continuously into kilometer-long range ordered macroscopic fibers (Fig. 1a and Supplementary Video 1). It is significant to study the rheological behavior of liquid crystal conjugated polymer during the spinning process, especially the isotropic and anisotropic phase equilibria and liquid-crystal transformation under the action of shear forces.

Finite element simulations were conducted to examine the fluid behavior of the $BBL_{99}$ (the viscosity-average molecular weight of BBL is 33 kDa, the number of repeating units is 99) methanesulfonic acid ($BBL_{99}$-MSA) solutions with varying concentrations and flow velocities in the spinneret region. The results show that increasing the flow rate and concentration of $BBL_{99}$-MSA solutions can lead to an increase in shear forces within the flow field (Fig. 1b and Supplementary Fig. 1-4 and Supplementary Table. 1). By calculating the fluid parameters, the Reynolds number (Re) is 0.22 ~ 8.80 (Supplementary Table 2), indicating that the fluid in the columnar region of the spinning needle is in a laminar state (the known range is < 2000). Using a 5 mg ml$^{-1}$ $BBL_{99}$-MSA solution as an example, increasing the flow rate has substantially enhanced the shear forces within the fluidic field. For a 5 mg ml$^{-1}$ $BBL_{99}$-MSA solution, the shear force distribution obtained through fluid simulations closely corresponds with the transmission light intensity and distribution observed while using in-situ polarized light microscopy under varying flow rates (Fig. 1c, d and Supplementary Fig. 5). At 45° to the direction of the analyzer, the fluid exhibits birefringence, and the intensity of transmitted light increases with increasing flow rate. The anisotropy observed is comparable to that in high-concentration BBL$_{99}$-MSA (90 mg ml$^{-1}$) liquid crystal solutions, suggesting that the liquid crystal phase of BBL at low concentrations is augmented due to the action of shear forces (Supplementary Fig. 6). It is particularly important to note that, to our knowledge, this is the first observation of the liquid crystalline phase texture of BBL in Lewis acids.

Furthermore, by regulating the polymer concentration, solutions of poly(benzimidazobenzophenanthroline) semiladder (BBB)-MSA, BBL$_{39}$ (the viscosity-average molecular weight of BBL is 13 kDa, the number of repeating units is 39)-MSA, poly(benzodifurandione)(PBFDO)-DMSO, and poly(3,4-ethylenedioxythiophene) polystyrene sulfonate (PEDOT:PSS)-H$_2$O have the same dynamic viscosity as the BBL$_{99}$-MSA solution. Thus, the fluid properties of conjugated polymer solutions under the same shear were characterized (Supplementary Table 3). Compared to the rigid rod-like molecule chain structure of BBL, BBB is a flexible chain polymer that does not form liquid crystalline mesophases in solution[19]. As a comparison supplement, no birefringence phenomenon was observed in the BBB-MSA solution (Supplementary Fig. 7). With the increase in shear force, the intensity of transmitted light and the anisotropy of molecular chains also increase in the PBFDO-DMSO and BBL$_{39}$-MSA solution (Supplementary Fig. 8-9). However, almost no change in light intensity was observed with PEDOT:PSS-H$_2$O (Supplementary Fig. 10), which may be related to its blend system and the molecular structure of its flexible main chain. Above all results indicate that the anisotropic phenomena under shear action are due to the formation of the liquid crystal phase of rigid rod-like conjugated polymers. Liquid crystal conjugated polymer BBL$_{99}$ fibers demonstrate a well-defined annular fibrous structure and can attain a radius of curvature of about 10 μm (Fig. 1e). The obtained microfibers showed diameters ranging from 16.10 ± 0.09, 17.06 ± 0.13 and 18.08 ± 0.11 μm, depending on the flow rate (Fig. 1f).

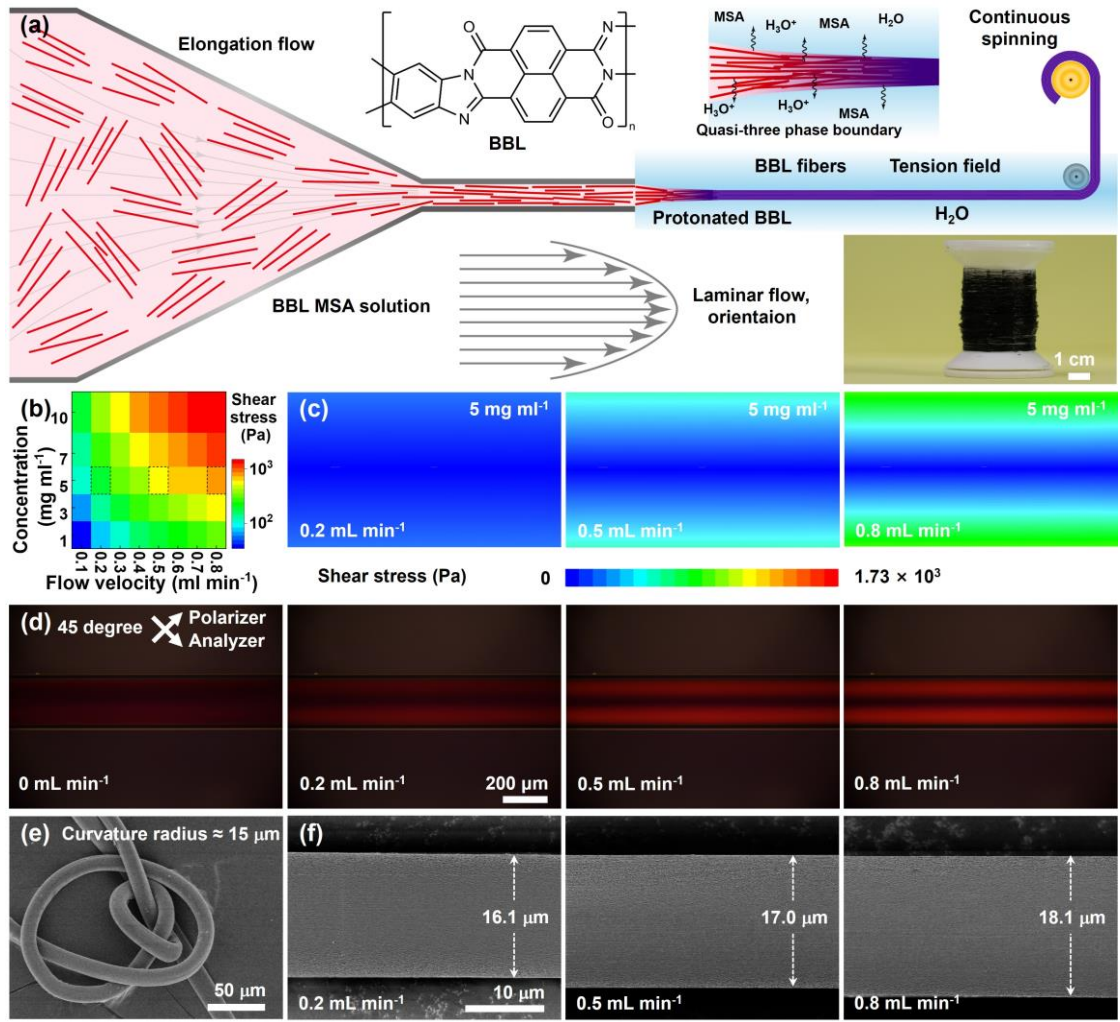

**Fig. 1. The continuous liquid crystal spinning of semiconductor fiber.** (a) Schematic fabrication, main transformation, and digital photograph of the macroscopic BBL semiconductor fibers. (b) Thermogram of the maximum shear rate for different flow velocities at different concentrations. (c) Finite element fluid simulation of the shear stress distribution of the flow field in a microfluidic channel, with a needle diameter of 200 μm, at spinning speeds of 0.2 (left), 0.5 (middle), and 0.8 (right) mL min$^{-1}$. (d) POM images of the BBL MSA solutions in a 200 um-diameter glass capillary in pressure-driven nematic flows at 45 ° to the analyzer direction with 0 (left), 0.2 (middle left), 0.5 (center right) and 0.8 (right) mL min$^{-1}$. (e) SEM picture of BBL fiber in the bent state. (f) SEM images of the BBL fiber-0.2 (left), BBL fiber-0.5 (middle), and BBL fiber-0.8 (right).

**Microstructures of semiconductor fibers.**

Grazing-incidence wide-angle X-ray scattering (GIWAXS) was employed to analyze BBL$_{99}$ fiber microstructures (Fig. 2a–c). All the BBL$_{99}$ fibers show a strong lamellar (100) diffraction peak at around $q_z = 0.76$ Å$^{-1}$ (d-spacing = 8.27 Å) and a strong π–π stacking (010) peak at about $q_{xy} = 1.87$ Å$^{-1}$ (d-spacing = 3.37 Å) (Fig. 2d, e). As the shear effect increases during spinning, the π–π stacking distance decreases from 3.52 Å for BBL$_{99}$ fiber-0.2 to 3.37 Å for BBL$_{99}$ fiber-0.8. The slight reduction in the π-π stacking distance suggests that the BBL polymer chains are more aligned under larger shear forces, resulting in stronger π-π interactions. (Figure 2f and Supplementary Fig. 11). The results of BBL$_{39}$ fiber also show the same trend, but the changes are relatively minor due to its shorter molecular chains (Figure 2g and Supplementary Fig. 12).

Microstructures of BBL$_{99}$ fiber-0.8 and films were compared using wide-angle X-ray scattering (WAXS) (Supplementary Fig. 13). Both BBL$_{99}$ fibers and films show a strong π–π stacking (010) peak at around $q_z$ = 1.87 Å$^{-1}$ (d-spacing = 3.37 Å), while only BBL fibers show a strong lamellar (100) diffraction peak at around $q_z$= 0.76 Å$^{-1}$ (d-spacing = 8.27 Å), which is consistent with the GIWAXS. WAXS diffractograms reveal that BBL fibers were oriented along the spinning direction via extrusion through the nozzle, as indicated by the reflection ring that evolved into arcs. The WAXS curve at the (100) peak indicates that the orientation of the crystalline region of the BBL$_{99}$ fiber-0.8 is approximately 67.14%. Compared to the isotropy exhibited by thin film materials, fibers prepared with shear-enhanced liquid crystals significantly improve orientation.

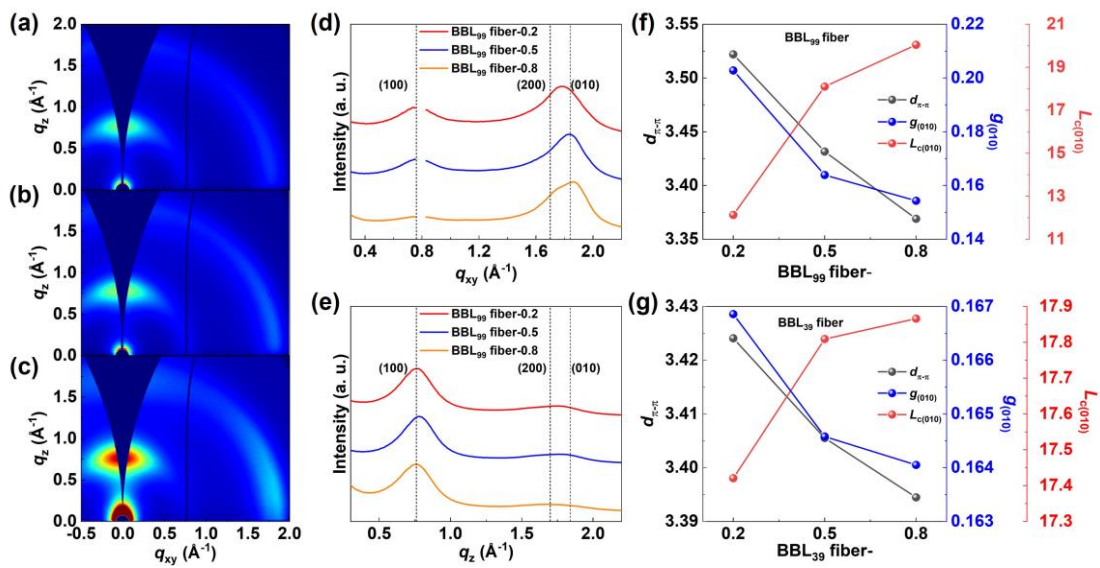

**Fig. 2. BBL fibers microstructures.** (a–c) 2D GIWAXS patterns of BBL$_{99}$ fiber-0.2 (a), BBL$_{99}$ fiber-0.5 (b), and BBL$_{99}$ fiber-0.8 (c). (d,e) In-plane (d) and out-of-plane (e) GIWAXS line cuts of BBL$_{99}$ fiber. (f) π–π distance ($d_{π–π}$), coherence length ($L_{c(010)}$), paracrystalline disorder ($g_{(010)}$) of BBL$_{99}$ fiber. (g) $d_{π–π}$, $L_{c(010)}$, and $g_{(010)}$ of BBL$_{39}$ fiber.

## Semiconductor fiber OECT

Organic Electrochemical Transistors (OECTs) have advantages such as low driving

voltage (<1 V), high transconductance (>1 mS), and good compatibility with logic circuits, giving fibrous OECT devices unique advantages in wearable electronics. So, we evaluated the impact of shear-enhanced liquid crystal phases on the functionality of $BBL_{99}$ fiber OECTs. Electrical/electrochemical properties of BBL fibers were extracted from OECT transfer/output measurements. The $BBL_{99}$ fiber was placed on the patterned source and drain surfaces, followed by covering the active channel with electrolyte (0.1 M NaCl) and submerging it in a silver chloride electrode as a gate bias (Fig. 3a). All $BBL_{99}$ fiber OECTs show typical n-type accumulation-mode output and transfer characteristics, as well as excellent reproducibility with an ON current standard deviation of less than 4.7% for five different devices (Supplementary Fig. 14). The $I_{ON}/I_{OFF}$ ratio increases from $2.5 \times 10^3$ for $BBL_{99}$ fiber-0.2 to $6.5 \times 10^3$ for $BBL_{99}$ fiber-0.8 OECTs (Fig. 3b, c and Supplementary Fig. 15). At $V_D = V_G = 0.7$ V, $I_D$ reaches the maximum value of $0.299 \pm 0.014$ mA for $BBL_{99}$ fiber-0.2 and increases up to $0.461 \pm 0.015$ mA for $BBL_{99}$ fiber-0.8. The maximum $g_{m,norm}$ increases from $2.219 \pm 0.009$ S cm$^{-1}$ for $BBL_{99}$ fiber-0.2 to $2.744 \pm 0.037$ S cm$^{-1}$ for $BBL_{99}$ fiber-0.8 (Fig. 3c and Supplementary Tables 4), with the latter being the highest transconductance reported to date for n-type fiber OECTs. We then calculated $\mu C^*$ to be $5.911 \pm 0.179$, $6.594 \pm 0.316$ and $7.659 \pm 0.484$ F cm$^{-1}$ V$^{-1}$ s$^{-1}$ for $BBL_{99}$ fiber-0.2, $BBL_{99}$ fiber-0.5, and $BBL_{99}$ fiber-0.8, respectively (Supplementary Tables 5). As $C^*$ does not change significantly within the series, it has average values in the 655 ~ 667 F cm$^{-3}$ (Supplementary Fig. 16-19). The carrier mobility ($\mu$) ranges from $(8.975 \pm 0.646) \times 10^{-3}$ cm$^2$ V$^{-1}$ s$^{-1}$ for $BBL_{99}$ fiber-0.2 to $(1.153 \pm 0.080) \times 10^{-2}$ cm$^2$ V$^{-1}$ s$^{-1}$ for $BBL_{99}$ fiber-0.8 (Figure 4d). The improved OECT properties of $BBL_{99}$ fibers are closely related to strong π-π interaction and high crystallinity, which is also consistent with the GIWAXS results indicating that the formation of solution-enhanced liquid crystal phases leads to enhanced charge transport properties of the rigid BBL polymer backbone[20,21]. The foremost demand when applying OECTs as a logic signal processing or sensor element is long-term stable operation without any deterioration of the equipment performance. Fig. 3e,f demonstrated the robustness of the electrical output yielded in a $BBL_{99}$ fiber OECT at $V_D = 0.6$ V with sequential gate bias pulses ($V_G = 0.6$ V, pulse duration = 7 s) for almost 1 h. The slight decrease in the $I_{ON}/I_{OFF}$ ratio as the test proceeds was probably due to the increase in the aqueous electrolyte concentration caused by the evaporation of solvent water from the channel. After adding 20 μL of electrolyte (0.1 mol L$^{-1}$ NaCl) again, the performance of BBL$_{99}$ fiber OECT recovered and still maintained excellent stability (Supplementary Fig. 20a). This phenomenon demonstrated the outstanding operational stability of BBL fibers in aqueous electrolytes, as also indicated by its remarkably high electron affinity, and were in close concordance with prior research. The response time of BBL$_{99}$ fiber OECT was assessed by exponential fitting to be 0.044 s for turning on (Supplementary Fig. 20b). The electrochemical characteristics of previously reported fiber-based OECTs are summarized in Supplementary Table 4.

According to the WAXS data, BBL fibers have a higher orientation than the film form, and charge transport in organic semiconductors strongly depends on molecular orientation. We evaluated the OECT performance of BBL$_{99}$ fibers along axial and radial directions. BBL$_{99}$ fiber-0.8 (diameter of about 100 μm) was placed in the wet state along the radial direction in the channel between the source and the drain. The remaining setup matched the description above, which constructed BBL$_{99}$ fiber radial OECTs (Fig. 3g). We then calculated $\mu C^*$, $\mu$, and maximum $g_{m,norm}$ to be 1.761 F cm$^{-1}$ V$^{-1}$ s$^{-1}$, 2.640 × 10$^{-3}$ cm$^2$ V$^{-1}$ s$^{-1}$, and 0.692 S cm$^{-1}$ for BBL fiber-radial OECTs, respectively (Supplementary Fig. 21). Benefiting from the predominant axial alignment of macromolecules in BBL$_{99}$ fibers, the axial $\mu C^*$, $\mu$ and maximum $g_{m,norm}$ exhibit a 436%, 436%, and 397% increase, respectively, compared to the radial direction.

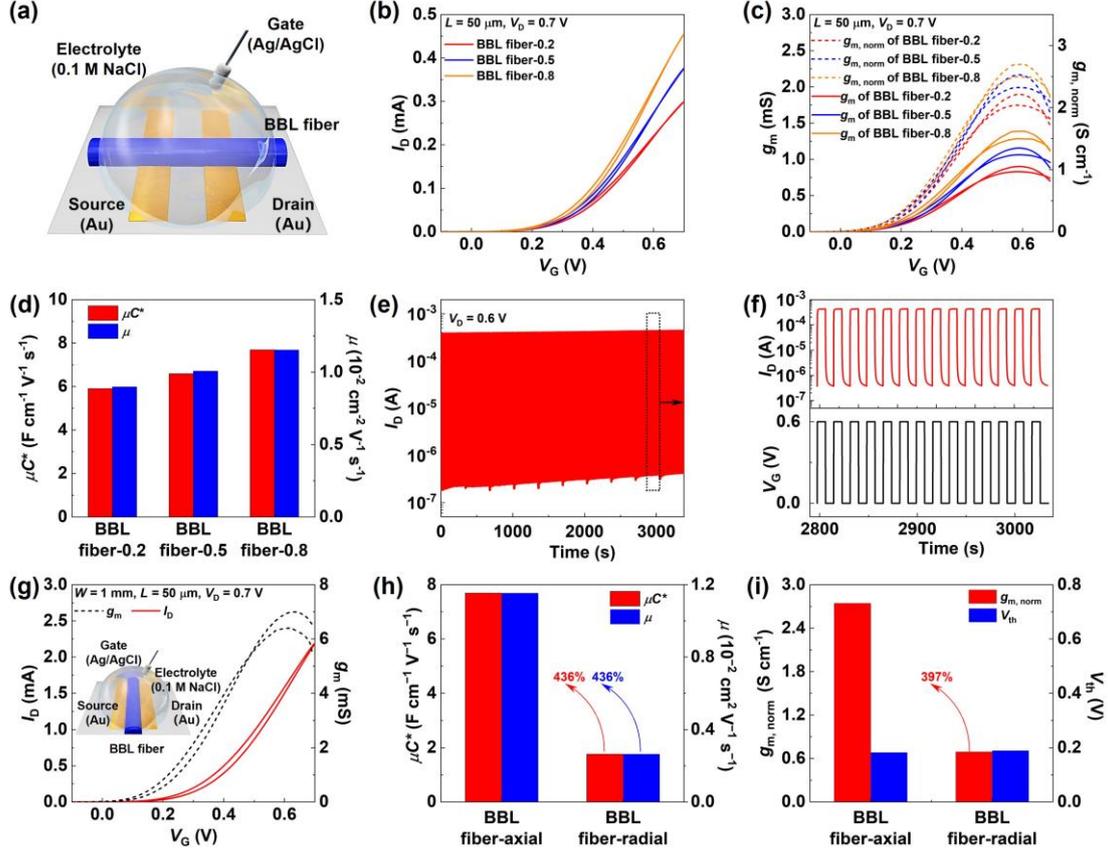

**Fig. 3. Semiconductor fiber OECT.** (a) Schematic diagram of BBL fiber-axial OECT. (b) Transfer curves of BBL$_{99}$ fiber. (c) Transconductance and geometry-normalized transconductance of BBL$_{99}$ fiber. (d) $\mu C^*$ and $\mu$ of BBL$_{99}$ fiber, derived from the OECT transfer characteristics. (e) Stability characterization of the BBL$_{99}$ fiber OECT at successive square wave gate bias for almost 1 h ($V_G$ = 0.6 V, pulse length = 7 s). (f) Amplification was performed for 4 min at t = 2920 s. All the above BBL$_{99}$ fiber axial OECTs have the same channel geometry ($L$ = 50 μm; the size of BBL fibers is defined as $A$). (g) Schematic diagram, transfer, and transconductance curve of BBL$_{99}$ fiber axial. (h) Comparison of $\mu C^*$ and $\mu$ for BBL$_{99}$ fiber axial and BBL$_{99}$ fiber radial OECT. (i) Comparison of $g_{m, norm}$ and $V_{th}$ for BBL$_{99}$ fiber axial and BBL$_{99}$ fiber radial OECT. All the above BBL$_{99}$ fiber radial OECTs has the same channel geometry ($L$ = 50 μm; $W$ = 1000 μm; $d$ = 5.06 μm).

**The alignment and mechanical performance of semiconductor fibers**

The BBL$_{99}$ fiber-0.8, for example, was further verified for alignment using polarized optical microscopy (POM), a recognized tool for fiber alignment characterization (Fig. 4a and Supplementary Fig. 22). It demonstrated that high orientation only at the edge of the fiber as polarized light is unable to pass through. The internal structure of the fiber was explored by small-angle X-ray scattering (SAXS). Perform 1D integration on the 2D SAXS scattergram and obtain the distribution curve of the scattering intensity with the value of $q$ (Supplementary Fig. 23). There is an obvious diffraction arc near the equatorial direction $q$ = 0.1435, but none in the meridian direction. This indicates that the BBL macromolecular chains in the fiber are highly oriented along the fiber axis so that the molecular chains are arranged in an orderly lateral direction to form a crystalline region. In contrast, perpendicular to the fiber axis, the longitudinal direction remains disordered.

Furthermore, the polarized FT-IR spectra of the $BBL_{99}$ fiber confirmed the existence of the alignment in the fiber, as reflected by the differences in the intensity of non-polarized, perpendicular ($\perp$), and parallel (//) polarized orientations (Supplementary Fig. 24) with perpendicular light showing the highest absorbance. The intensity differences stem from the polymer chain orientation, reflected in the varying quantities of each functional group aligned with each electric field vector. Strong absorption at 1000 cm$^{-1}$ was indicative of the majority of trans-BBL conformers. The C=O vibration in BBL is located at 1705 cm$^{-1}$, which should be perpendicular to the polymer backbone. The dichroic ratio of this band was $R = 0.52$, which indicated higher absorbance with perpendicular polarization. Since the direction of the BBL molecular chain and C=O bond is 65.89°, the calculated orientation factor ($f$) is 0.76. π-π stacking is a type of dispersion effect (van der Waals) and is crucial in the total intermolecular energy related to the packing of polymer chains[22,23]. BBL molecules have planar rigid backbones that are stacked in a face-to-face manner, resulting in superior intermolecular π-π interaction strength between polymer chains.

As the interaction between BBL molecules strengthens, the tensile strength of $BBL_{99}$ fibers increases from 217.5 MPa for $BBL_{99}$ fiber-0.2 to 337.2 MPa for $BBL_{99}$ fiber-0.8. The mechanical properties of polymer fibers also depend closely on molecular weight. The fiber spun from high-molecular-weight $BBL_{165}$ (the viscosity-average molecular weight is 54.9 kDa, the number of repeating units is 165) exhibited a tensile strength of 595.1 MPa at 54% elongation and a Young's modulus of 15.7 GPa. It is significantly higher than the tensile strength of fibers used for commercial fabrics and other semiconductor fibers[24–26], endowing it with the capacity for industrial-scale manufacturing (Fig. 4c and Supplementary Fig. 25). The UV resistance, chemical stability, and thermodynamic stability of fibers are also crucial for application to intelligent textiles (Fig. 4d and Supplementary Fig. 26). To explore the UV resistance of $BBL_{99}$ fiber, after exposure to UV irradiation (5000 W m$^{-2}$ 365 nm UV, about 3500 times the conventional UV light intensity) for 12 h, the tensile strength of the $BBL_{99}$ fibers was still maintained at 96.28%. After being soaked in acetone (volume concentration, 99.5%), NaOH (mass concentration, 30%), and $H_2SO_4$ (volume concentration, 90%) solutions for 24 h, respectively, $BBL_{99}$ fibers maintained clear filamentous shapes. The tensile strength was maintained at 94.03%, 91.26%, and 90.31% showing excellent chemical stability.

Additionally, the high crystallinity and orientation caused by shear-enhanced liquid crystal phases ensure that BBL semiconductor fibers maintain high mechanical strength and stability even in extreme temperatures. Thermogravimetric analysis (TGA) indicated that the mass of $BBL_{99}$ fibers was preserved at about 98.2% under 500°C, which also proved the excellent high-temperature thermal stability of the $BBL_{99}$ fibers (Supplementary Fig. 27a). The dynamic mechanical analysis (DMA) results of $BBL_{99}$ fibers show outstanding thermomechanical stability (Fig. 4e). $BBL_{99}$ fibers can remain stable even in liquid nitrogen (Fig. 4f and Supplementary Fig. 27b-d and Supplementary video 2). According to the Grüneisen vibrational theory of thermal expansion, a negative volumetric thermal expansion coefficient is due to the negative Grüneisen parameter that implies a decrease in the phonon frequency (softening of the acoustic phonon modes) and the highly aligned and crystallized structures of $BBL_{99}$ fibers[27,28]. Compared to PEDOT:PSS fibers (Supplementary Fig. 28, +2.07%), $BBL_{99}$ fibers have a lower mechanical strain and a negative coefficient of thermal expansion of about -1.08% in strain. The above results demonstrate the excellent mechanical properties and stability of solution-enhanced liquid crystal BBL fibers in various environments.

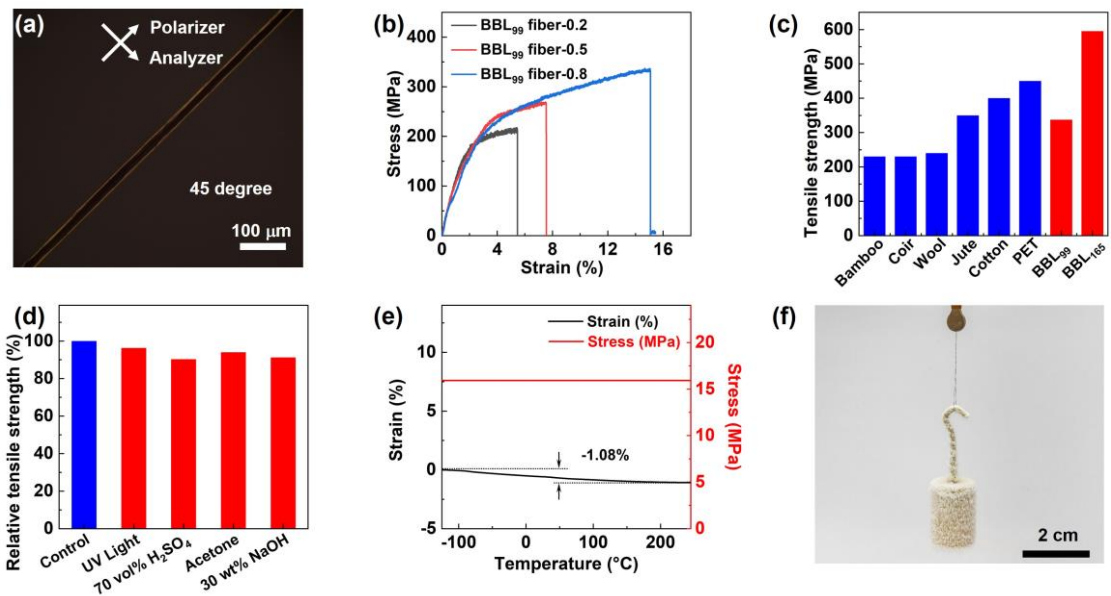

**Fig. 4. The alignment and mechanical performance of BBL fibers.** (a) The POM image of the BBL$_{99}$ fiber-0.8 when the fiber axis was at 45° to the analyzer direction. (b) Tensile strength of the BBL$_{99}$ fiber. (c) Tensile strength of the BBL$_{99}$ fiber-0.8 (337.2 MPa), BBL$_{165}$ fiber-0.8 (595.1 MPa), and other representative fiber used in commercialized clothes. (d) Relative tensile strength of BBL$_{99}$ fibers-0.8 after UV radiation and chemical reagent treatment, respectively. (e) DMA graph of BBL$_{99}$ fiber-0.8 (40 um) under 16 MPa; (f) Digital photograph of BBL$_{165}$ fiber bundle (20 um diameter) hanging with a 10 g weight in liquid nitrogen after 3 minutes.

**Fabric level logic circuits**

The NAND gate is the most important logic gate in digital electronics, which can be used to build combinational logic circuits such as adders, data selectors, etc. After successfully demonstrating high-performing n-type accumulation-mode fiber OECTs, we continued fabricating NAND logic circuits with 2, 3, 4, and 5 input signals (Fig. 5a-c and Supplementary Fig. 29). Based on the above results, semiconductor fibers are promising candidates for fabricating large-scale 0-1 logic gate circuits, thus enabling real-time portable logic operations. Meanwhile, given the remarkable electrochemical and mechanical properties described above, BBL fibers can be widely applied in the fabrication of fabric logic circuits. To better integrate information conversion, digital, and logical calculations into smart clothing and move closer to the ultimate dream of "fabric chips", we have used weaving technology to realize the construction of fabric-level logic circuits using semiconductor fibers. BBL fibers are combined onto polyimide fibers with patterned electrodes on the surface to construct source, drain, and channel parts. Furthermore, Ag/AgCl coating is applied to surface conductive nylon fibers as a gate. The functional fibers mentioned above were woven into a fabric, and the fabric-level OECTs were fabricated by applying solid electrolytes at the fabric joints. Acting as an electronic switch in a circuit is the fundamental and critical role of a transistor device. The fabric BBL OECT is connected in series with the LED drivers and woven into the fabric (Fig. 5d-e). As the $V_G$ of the fabric OECT increased, so did the current in the circuit and the brightness of the RYB LED bulb increased until $V_G$ = 0.6 V. At that point, the OECT-LED circuit is in a fully conductive state (Fig. 5f). Fabric NAND gate logic operation circuits were also fabricated using BBL fabric OECTs (Fig. 5g-h). The fabric NAND gate circuits exhibit the same excellent performance as planar devices and can communicate digital signals such as ON/OFF (Fig. 5i). ON-to-OFF output transformations (digital signal "0" to digital signal "1") enable correct logic functions. Based on this, more complex structured logic devices and circuits can be manufactured, which reveals the prospect of intelligent integrated circuits for generating sensible logic judgments.

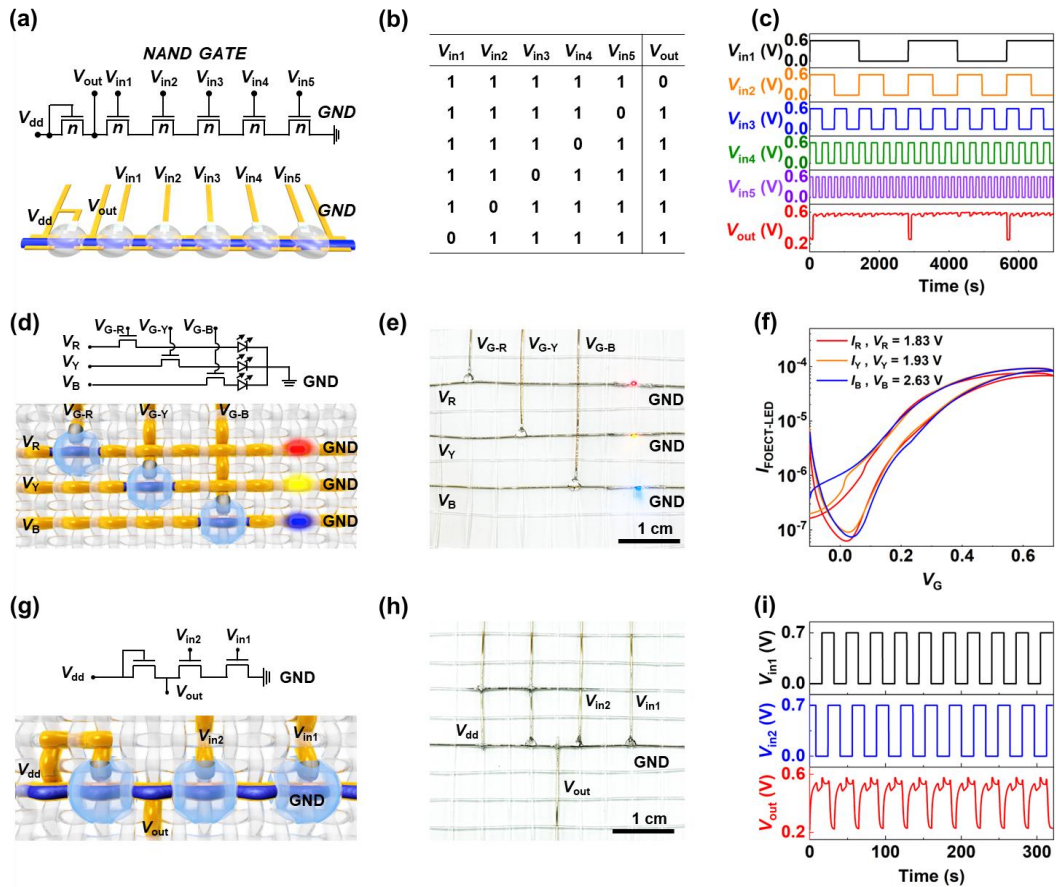

**Fig. 5. BBL fiber logic circuits to fabric level manufacture.** (a) A schematic diagram of the single fiber

NAND logic circuits of 5 input signals. (b) The truth table of the single fiber NAND logic circuits of 5 input signals. (c) Output characteristics of BBL fiber NAND logic circuits. (d) A Schematic and circuit diagram using fabric BBL OECT as a circuit switch in series with RYB LEDs. (e) A photograph of the fabric OECT-LED circuits. (c) The variation of positive volt-ampere characteristics of fabric circuits. (g) A Schematic and circuit diagram of fabric NAND gate circuit of 2 input signals. (h) A photograph of the fabric NAND gate. (i) Output characteristics of fabric NAND gate at $V_{dd}$ = 0.8 V. Logic '0' and '1' mean 0.23 V and 0.57 V, respectively. The above devices were prepared based on BBL$_{99}$ fiber-0.8.

## Conclusions

In summary, we investigated the control of π-π stacking interactions and orientation in BBL molecules under fluid shear enhanced by liquid crystals, thus developing a continuous spinning strategy that simultaneously improves the mechanical strength and electrochemical performance of conjugated polymers. The results demonstrate that the inherent orientation of liquid crystal molecules, coupled with secondary orientation due to fluid shear, imparts BBL fibers with a pronounced axial preferred orientation, leading to anisotropic electrochemical performance. The axial carrier mobility and transconductance increased by ~400% in the radial direction. The fabricated semiconductor fibers feature mechanical properties (600 MPa) and environmental stability suitable for industrial garment manufacturing, showcasing their potential for practical logic textile production. Finally, we demonstrated a NAND logic circuit based on semiconductor fiber OECTs. This highlighted the new possibilities for developing high-performance soft-fiber electronic devices and their immense potential for fabric-level logic chip applications.

## Methods

### Materials

BBL was synthesized based on previous studies[29] with viscosity-average molecular weights ($M$v) of 13kDa, 32.4 kDa, and 54.9 kDa, named as BBL$_{39}$, BBL$_{99}$, and BBL$_{165}$. BBB was synthesized based on previous studies[30]. Methanesulfonic acid (MSA), NaOH (97%), and ethanol were purchased from Aladdin. Poly (sodium 4-styrenesulfonate) (average $M$w ≈ 1000 kDa, powder) was obtained from Sigma Aldrich. Concentrated sulfuric acid (≈98%), glycerol (≥99.0%), D-sorbitol (≥99.5%), NaCl (≥99.5%), and acetone (AR, ≥99.5%) were obtained from China National Pharmaceutical Group Corporation. Unless otherwise mentioned, all materials were utilized as they were.

**Semiconductor fiber preparation**

The $BBL_{99}$ fiber was fabricated through a liquid crystal wet spinning method. $BBL_{99}$ was dissolved in MSA (5 mg ml$^{-1}$) under highspeed stirring. The $BBL_{99}$-MSA Spinning dope was pumped into the deionized water-filled coagulation bath at 20 °C with a rate of 0.2, 0.5, and 0.8 mL min$^{-1}$ using a syringe thruster (KD Scientific, Legato 101) and different diameters of spinning needles. During the solvent exchange, dark purple BBL fibers were generated. The coagulated BBL nascent fibers were drawn for the first stage of drafting and the second stage of drafting, the second acetone-filled coagulation bath, and the second stage of drafting while stretched to a draw ratio of 1:1.2. Finally, the fibers were continuously collected on fiber spools for drying and then annealed at 200°C. Except that the concentration of the $BBL_{39}$-MSA solution is 10.1mg ml$^{-1}$, the preparation method for $BBL_{39}$ fibers is the same as for $BBL_{99}$ fibers. PBFDO-DMSO solution (10-12 mg ml$^{-1}$, VOLT-AMP Optoelectronics Tech. Co.) was diluted to 7.6 mg ml$^{-1}$ as the spinning solution, with $H_2SO_4$-$H_2O$ solution used as the coagulation bath, spun with the same parameters as $BBL_{99}$, and dried at room temperature to obtain PBFDO fibers. PEDOT:PSS aqueous solution (1.1–1.3% solid content, Clevios™ PH1000, Heraeus Electronic Materials) was freeze-dried to solid PEDOT:PSS and then configured to 6.8 mg ml$^{-1}$ aqueous dope. PEDOT:PSS fibers were obtained by using $H_2SO_4$ solution as the coagulation bath and spinning with the same parameters as $BBL_{99}$. The fibers were dried at room temperature.

**Fluid Simulation and Finite Element Analysis**

Each conservation equation in this work was solved using a phase-coupled SIMPLE (PC-SIMPLE) method built on a finite space framework in C++. Up to 100 iterations were performed in the fluid simulation model to avoid remnant values of momentum, fluid, volume fraction, and other scalars, which ensured the convergence limit was below 10$^{-5}$. In particular, the enthalpy conservation equation was lower than 10$^{-7}$.

**Polarized optical microscopy and in-situ**

Polarized optical microscopy (POM) images of the BBL fibers were obtained with Olympus BX60 optical microscope. The fluid flow in the microchannel is driven and precisely controlled by using a syringe pump system (KD Scientific, Legato 101). $BBL_{99}$-MSA were configured as experimental fluids at a concentration of 5 mg ml$^{-1}$. Glass capillaries with inner diameters of 200 μm is used as microfluidic channels for in-situ polarization analysis to characterize the fluid state during the spinning process.

**Fourier transform infrared spectroscopy**

Fourier Transform Infrared Spectroscopy (FTIR) spectrum of BBL fibers was acquired in transmission mode inside a hermetic sample chamber with a Bruker Equinox 55 spectrometer.

**X-ray Scattering**

$BBL_{99}$-fiber: Grazing-incidence wide-angle X-ray Scattering (GIWAXS) experiments were performed at Xenocs Xeuss 2.0. The X-ray energy was 8.73 keV, and the incidence angle was 0.12°. The total exposure time was 1800 s. The scattered X-rays were recorded by a charge-coupled device detector located 210.783 mm from the sample. $BBL_{39}$-fiber: GIWAXS experiments were performed at the beamline BL02U2 system at Shanghai Synchrotron Radiation Facility (SSRF), China. The X-ray energy was 9.5 keV and the incidence angle was 0.12°. The total exposure time was 10 s. The scattered X-rays were recorded by a charge-coupled device detector located 207.456 mm from the sample.

Wide Angle X-ray Scattering (WAXS) experiments were measured with a Rigaku Rapid II X-ray, and the wavelength of the X-ray source was 0.154 nm. The axial direction of BBL fibers is perpendicular to the incident X-ray beam and parallel to the meridian of the two-dimensional WAXS pattern. The BBL fibers and film samples were exposed for 15 s and 60s at ambient conditions. The scattering images were captured with a Hypix-6000 detector; the sample-to-detector distance was 70 mm.

Small-angle X-ray scattering (SAXS) measurements were undertaken on the beamline BL19U2 SAXS system at SSRF, China. The radiation energy of the X-rays used was 12.4 keV at the wavelength of 1.03 Å. The distance between the detector and the fibers is 3180 mm. The BBL fibers were placed in neatly oriented bundles, anchored parallel to a suitable sample stage with an aperture for X-ray irradiation. Samples exposure to X-rays for 3 s to obtain 2D SAXS patterns of the BBL fibers.

**Morphology and mechanical properties of fiber**

The surface and cross-sectional scanning and transmission electron microscopy (SEM) images of fibers were obtained by FESEM (SU-8010, Hitachi, Japan). The mechanical properties of the fibers were investigated with the Instron universal material testing system (INSTRON 5969, USA). The stretch rate was 100 mm·min$^{-1}$.

**Thermomechanical properties of fiber**

Thermomechanical properties were investigated in a constant force mode utilizing a dynamic mechanical analysis (DMA) (DMA Q800 TA Instruments). The fiber was held under a controlled force (0.02 N for 40 μm fibers, 0.005 N for 20 μm fibers), and the strain was monitored during a temperature ramp (5 K min$^{-1}$) from -140 to 300°C. Characterization performed at the milligram level provides properties on the chemical reactivity of solids materials with temperature to investigate the thermal stability of BBL fibers. Thermogravimetric analysis (TGA) was carried out using TGA/DSC 3+ Mettler Toledo and DSC 214 Polyma Netzsch. The loss rate and heat flow of solid materials were obtained in nitrogen at 10 K min$^{-1}$.

**Environmental stability of fibers**

Ultraviolet (UV): The BBL$_{99}$ fibers were arranged in parallel on a glass slide and irradiated with a 365 nm UV LED lamp at a power of 5000 W m$^{-2}$ (about 3500 times the conventional UV light intensity) for 12 hours, and then the tensile test was performed. BBL fibers were handled in a constant-temperature humidifier sample chamber at 25°C.

Chemical reagent: A quantity of the fibers was positioned in acetone (volume concentration, 99.5%), H$_2$SO$_4$ (volume concentration, 70%), and NaOH (mass concentration, 30%) solutions. After 24 h, the processed fibers were cleaned with deionized water and dried naturally at room temperature for tensile characterization.

**Fiber OECT and NAND logic circuit fabrication and characterization**

The metals employed for the electrodes were 3 nm Cr and 30 nm Au. The width of the transistor channel was 50 μm, and the length depended on the fiber diameter. 0.1 mol ml$^{-1}$ NaCl aqueous solution was applied as the electrolyte, and an Ag/AgCl pellet electrode was employed as the gate electrode.

Electrochemical impedance spectroscopy (EIS) characterizations were carried out on the Princeton Electrochemical Workstation (PMC CHS08A) at room temperature. For EIS characterizations, 0.1 mol L$^{-1}$ NaCl aqueous solution was applied as the electrolyte, a platinum mesh (2.5 * 2.5 cm) served as the counter electrode, and Ag/AgCl (saturated) was employed as the reference electrode. For volumetric capacitance characterization, WE were set to −0.7 V, and spectra were measured at $10^{-1}$ Hz–$10^2$ Hz. The results were fitted with employing the equivalent circuit model *Rs(Rp||Q)*.

OECTs and inverters were performed utilizing a Keithley 4200 semiconductor parameter analyzer. The continuous gate pulses were employed for the stability measurements through a Tektronix 3390 wave function generator.

**Fabric-level logic circuit**

Gel electrolyte: The PSSNa, glycerol, D-sorbitol, and deionized water (weight ratio of 5: 2: 2: 1) were mixed through a sonication and stirring process (50°C). The prepared gel electrolyte had pleasurable shape retention and perfect electrochemical performance.

Preparation of patterned electrodes on polyimide fibers by mask evaporation method and preparation of BBL composite fibers by laminating BBL fibers on the electrodes. The BBL composite fibers and fibers coated with Ag/AgCl electrodes and gel electrolyte were woven into the fabric. Dripping gel electrolytes prepared the fabric logic OECTs at the channel.

**Acknowledgments**

We are grateful for the National Natural Science Foundation of China (92163132, 52373282), the Science and Technology Commission of Shanghai Municipality (21520710700, 23ZR1402000), the Fundamental Research Funds for the Central Universities (2232021G-02, 2232021G-12). We are grateful to Shanghai Synchrotron Radiation Facility (SSRF) for providing the beam time under the project number 2023-NFPS-PT-502560 and 2023-NFPS-PT-500794.

**Data availability**

The authors declare that the primary data supporting the findings of this study are available within the article and its Supplementary Information. Additional data are available from the corresponding author (G.W.) upon request.

# Shear-enhanced Liquid Crystal Spinning of Conjugated polymer Semiconductor Fibers

# SUPPLEMENTARY MATERIAL

**Contents**

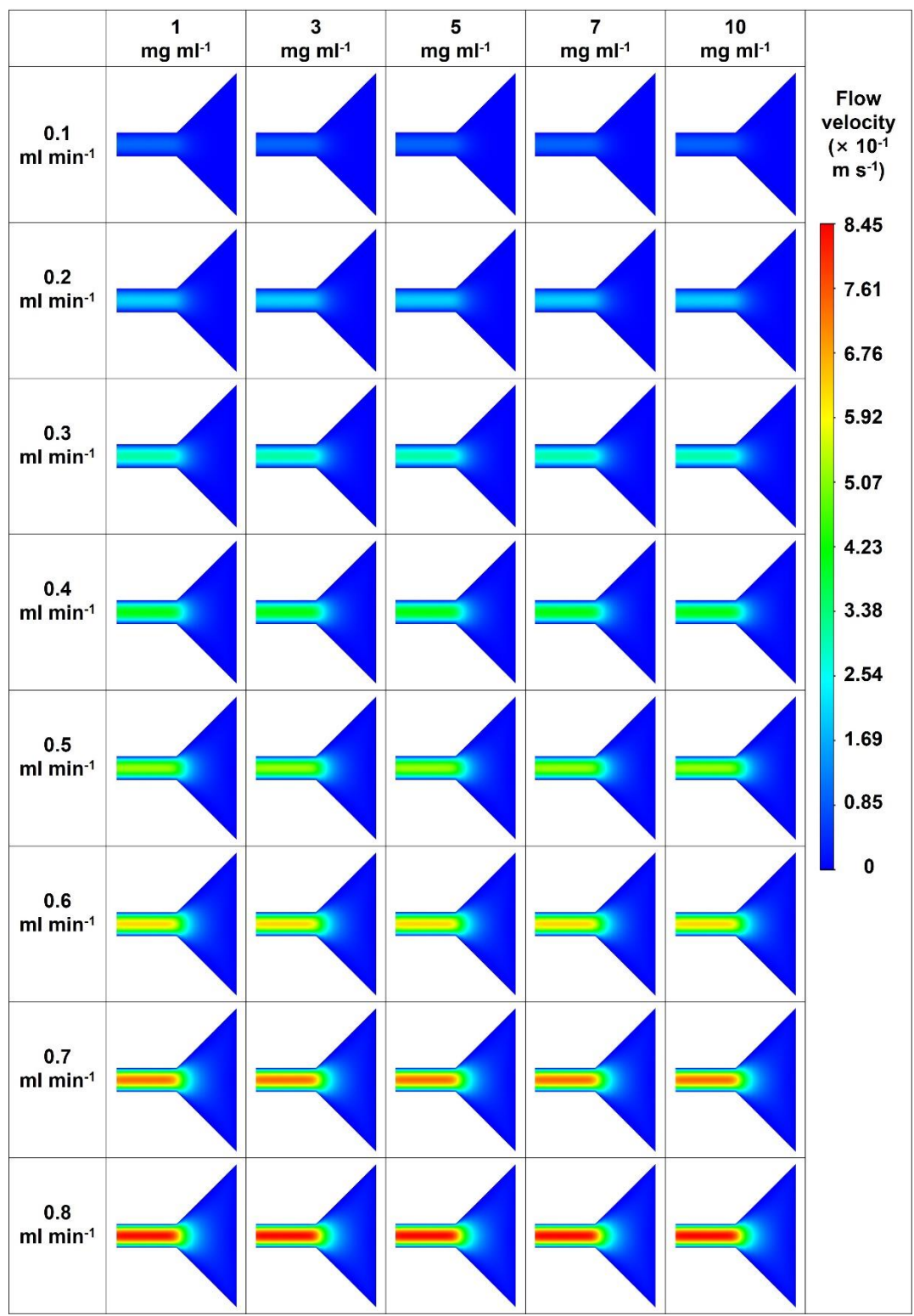

Supplementary Fig. 1 Finite element simulation of the flow velocity distribution. Finite element simulation of the flow velocity distribution in a 200 μm diameter microfluidic channel for different concentrations (1, 3, 5, 7, and 10 mg ml$^{-1}$) and different spinning speeds (0.1 ~ 0.8 ml min$^{-1}$) of BBL$_{99}$-MSA solution.

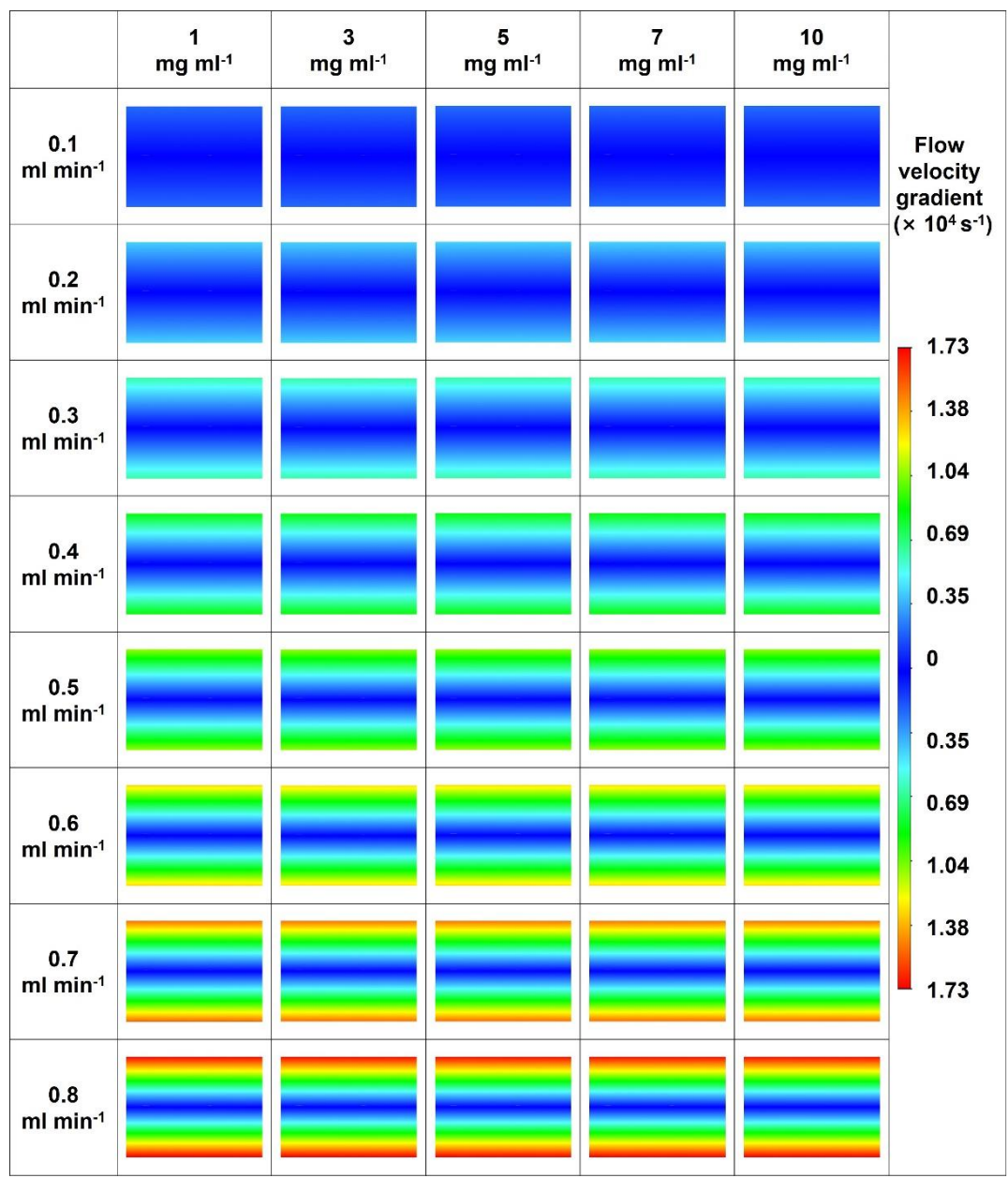

Supplementary Fig. 2 Finite element simulation of the flow velocity gradient distribution.

Finite element simulation of the flow velocity gradient distribution in a 200 μm diameter microfluidic channel for different concentrations (1, 3, 5, 7, and 10 mg ml$^{-1}$) and different spinning speeds (0.1 ~ 0.8 ml min$^{-1}$) of BBL$_{99}$-MSA solution.

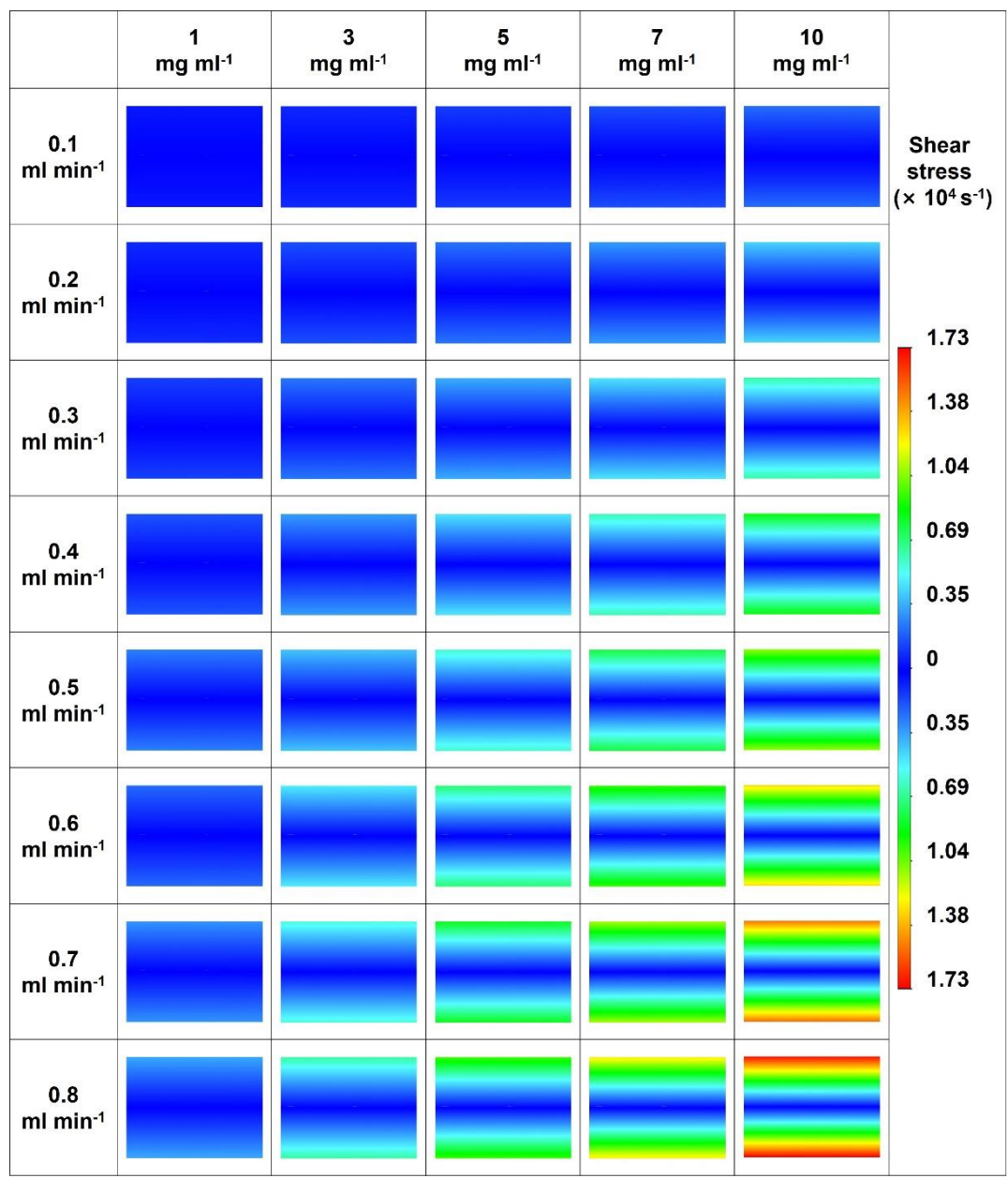

Supplementary Fig. 3 Finite element simulation of the shear stress distribution. Finite element simulation of the shear stress distribution in a 200 μm diameter microfluidic channel for different concentrations (1, 3, 5, 7, and 10 mg ml$^{-1}$) and different spinning speeds (0.1 ~ 0.8 ml min$^{-1}$) of BBL$_{99}$-MSA solution.

Supplementary Table 1 Dynamic viscosity and shear stress of BBL$_{99}$-MSA solutions.

| Concentration | Flow velocity | Density | Dynamic viscosity | Maximum shear stress |
|---|---|---|---|---|
| $c$ | $v$ | $\rho$ | $\mu$ | $\tau$ |
| mg ml$^{-1}$ | ml min$^{-1}$ | kg m$^{-3}$ | $\times 10^{-2}$ Pa s | pa |
| 1 | 0.1 | 1481 | 1.73 | 36.81 |
| 3 | 0.1 | 1483 | 3.23 | 69.25 |
| 5 | 0.1 | 1485 | 4.74 | 102.52 |
| 7 | 0.1 | 1487 | 6.25 | 136.48 |
| 10 | 0.1 | 1490 | 8.52 | 187.88 |
| 1 | 0.2 | 1481 | 1.73 | 73.61 |
| 3 | 0.2 | 1483 | 3.23 | 137.66 |
| 5 | 0.2 | 1485 | 4.74 | 202.26 |
| 7 | 0.2 | 1487 | 6.25 | 267.58 |
| 10 | 0.2 | 1490 | 8.52 | 367.53 |
| 1 | 0.3 | 1481 | 1.73 | 110.58 |
| 3 | 0.3 | 1483 | 3.23 | 206.45 |
| 5 | 0.3 | 1485 | 4.74 | 302.82 |
| 7 | 0.3 | 1487 | 6.25 | 399.67 |
| 10 | 0.3 | 1490 | 8.52 | 546.72 |
| 1 | 0.4 | 1481 | 1.73 | 147.70 |
| 3 | 0.4 | 1483 | 3.23 | 275.43 |
| 5 | 0.4 | 1485 | 4.74 | 403.67 |
| 7 | 0.4 | 1487 | 6.25 | 532.33 |
| 10 | 0.4 | 1490 | 8.52 | 727.03 |
| 1 | 0.5 | 1481 | 1.73 | 184.99 |
| 3 | 0.5 | 1483 | 3.23 | 344.54 |
| 5 | 0.5 | 1485 | 4.74 | 504.74 |
| 7 | 0.5 | 1487 | 6.25 | 665.24 |
| 10 | 0.5 | 1490 | 8.52 | 907.86 |
| 1 | 0.6 | 1481 | 1.73 | 222.43 |
| 3 | 0.6 | 1483 | 3.23 | 413.82 |
| 5 | 0.6 | 1485 | 4.74 | 605.96 |
| 7 | 0.6 | 1487 | 6.25 | 798.43 |
| 10 | 0.6 | 1490 | 8.52 | 1089.09 |
| 1 | 0.7 | 1481 | 1.73 | 260.02 |
| 3 | 0.7 | 1483 | 3.23 | 483.27 |
| 5 | 0.7 | 1485 | 4.74 | 707.31 |
| 7 | 0.7 | 1487 | 6.25 | 931.73 |
| 10 | 0.7 | 1490 | 8.52 | 1270.55 |
| 1 | 0.8 | 1481 | 1.73 | 297.83 |
| 3 | 0.8 | 1483 | 3.23 | 552.84 |
| 5 | 0.8 | 1485 | 4.74 | 808.81 |
| 7 | 0.8 | 1487 | 6.25 | 1065.22 |
| 10 | 0.8 | 1490 | 8.52 | 1452.21 |

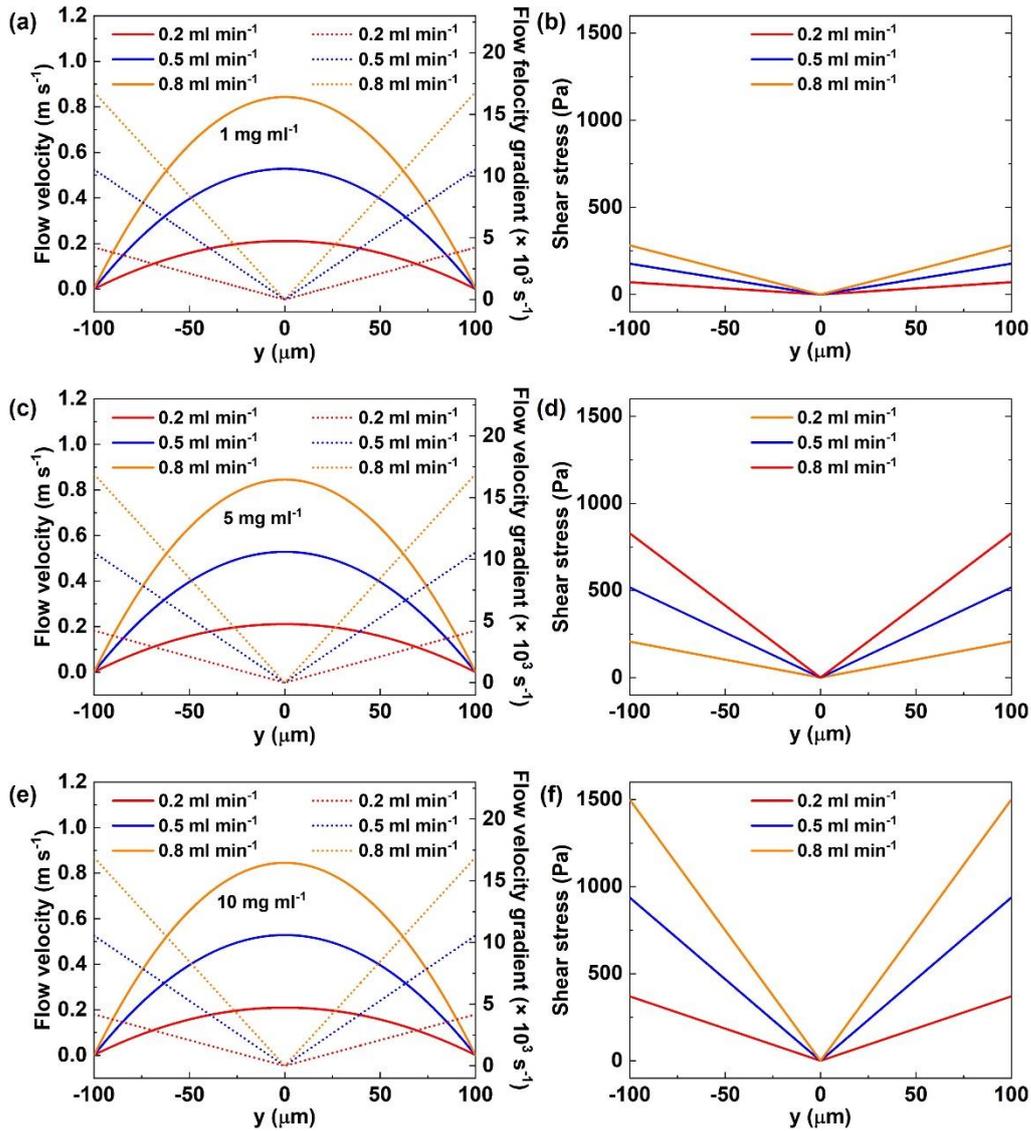

Supplementary Fig. 4 The flow velocity, flow velocity gradient, and shear stress of BBL$_{99}$-MSA solution.

The flow velocity, flow velocity gradient, and shear stress in a 200 μm diameter microfluidic channel for different concentrations (1, 3, 5, 7, and 10 mg ml$^{-1}$) and different spinning speeds (0.1 ~ 0.8 ml min$^{-1}$) of BBL$_{99}$-MSA solution.

Supplementary Table 2 Average velocities and Reynolds number of BBL$_{99}$-MSA solutions.

| Concentration | Flow velocity | Density | Average velocity | Reynolds number |
|---|---|---|---|---|
| $c$ | $v$ | $\rho$ | $v_{avg}$ | $Re$ |
| mg ml$^{-1}$ | ml min$^{-1}$ | kg m$^{-3}$ | $\times 10^{-1}$ m s$^{-1}$ | a. u. |
| 1 | 0.1 | 1481 | 0.064 | 1.090 |
| 3 | | 1483 | 0.064 | 0.587 |
| 5 | | 1485 | 0.064 | 0.400 |
| 7 | | 1487 | 0.064 | 0.303 |
| 10 | | 1490 | 0.064 | 0.222 |
| 1 | 0.2 | 1481 | 0.128 | 2.192 |
| 3 | | 1483 | 0.128 | 1.175 |
| 5 | | 1485 | 0.128 | 0.803 |
| 7 | | 1487 | 0.128 | 0.609 |
| 10 | | 1490 | 0.128 | 0.447 |
| 1 | 0.3 | 1481 | 0.192 | 3.294 |
| 3 | | 1483 | 0.192 | 1.763 |
| 5 | | 1485 | 0.192 | 1.205 |
| 7 | | 1487 | 0.192 | 0.915 |
| 10 | | 1490 | 0.192 | 0.672 |
| 1 | 0.4 | 1481 | 0.256 | 4.395 |
| 3 | | 1483 | 0.256 | 2.351 |
| 5 | | 1485 | 0.256 | 1.608 |
| 7 | | 1487 | 0.256 | 1.221 |
| 10 | | 1490 | 0.256 | 0.896 |
| 1 | 0.5 | 1481 | 0.320 | 5.496 |
| 3 | | 1483 | 0.320 | 2.938 |
| 5 | | 1485 | 0.321 | 2.010 |
| 7 | | 1487 | 0.320 | 1.526 |
| 10 | | 1490 | 0.320 | 1.121 |
| 1 | 0.6 | 1481 | 0.385 | 6.596 |
| 3 | | 1483 | 0.384 | 3.526 |
| 5 | | 1485 | 0.385 | 2.411 |
| 7 | | 1487 | 0.385 | 1.831 |
| 10 | | 1490 | 0.385 | 1.345 |
| 1 | 0.7 | 1481 | 0.449 | 7.696 |
| 3 | | 1483 | 0.448 | 4.113 |
| 5 | | 1485 | 0.449 | 2.813 |
| 7 | | 1487 | 0.449 | 2.137 |
| 10 | | 1490 | 0.449 | 1.569 |
| 1 | 0.8 | 1481 | 0.513 | 8.796 |
| 3 | | 1483 | 0.512 | 4.700 |
| 5 | | 1485 | 0.513 | 3.215 |
| 7 | | 1487 | 0.513 | 2.442 |
| 10 | | 1490 | 0.513 | 1.794 |

Supplementary Table 3 Concentration of the polymer and dynamic viscosity of the solution.

| Solute | Solvent | Concentration ($c$) mg ml$^{-1}$ | Density of the solution ($\rho$) kg/m$^3$ | Kinematic viscosity ($n$) mm$^2$/s | Dynamic viscosity ($m$) × 10$^{-2}$ Pa s |
|---|---|---|---|---|---|
| BBL$_{99}$ | MSA | 5 | 1485 | 31.9 | 4.737 |
| BBL$_{39}$ | MSA | 10.1 | 1490 | 31.5 | 4.694 |
| BBB | MSA | 24.1 | 1504.1 | 31.4 | 4.725 |
| PBFDO | DMSO | 7.6 | 1106.9 | 42.8 | 4.734 |
| PEDOT:PSS | H$_2$O | 6.8 | 1006.8 | 47.5 | 4.783 |

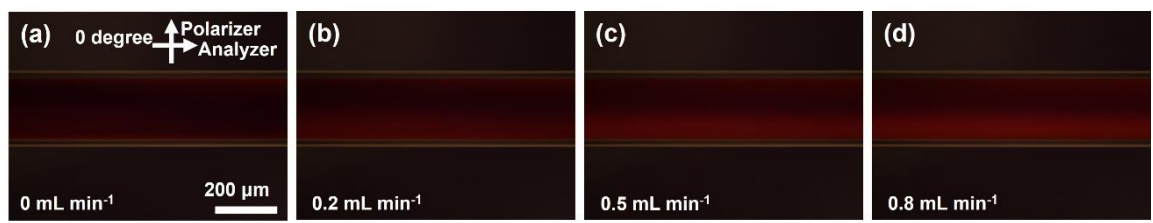

Supplementary Fig. 5 POM photographs of the BBL$_{99}$-MSA solutions.

POM photographs of the BBL$_{99}$-MSA solutions in a 200 μm-diameter glass tube in pressure-driven nematic flows at 0° to the analyzer direction with 0 (a), 0.2 (b), 0.5 (c), 0.8 (d) mL min$^{-1}$.

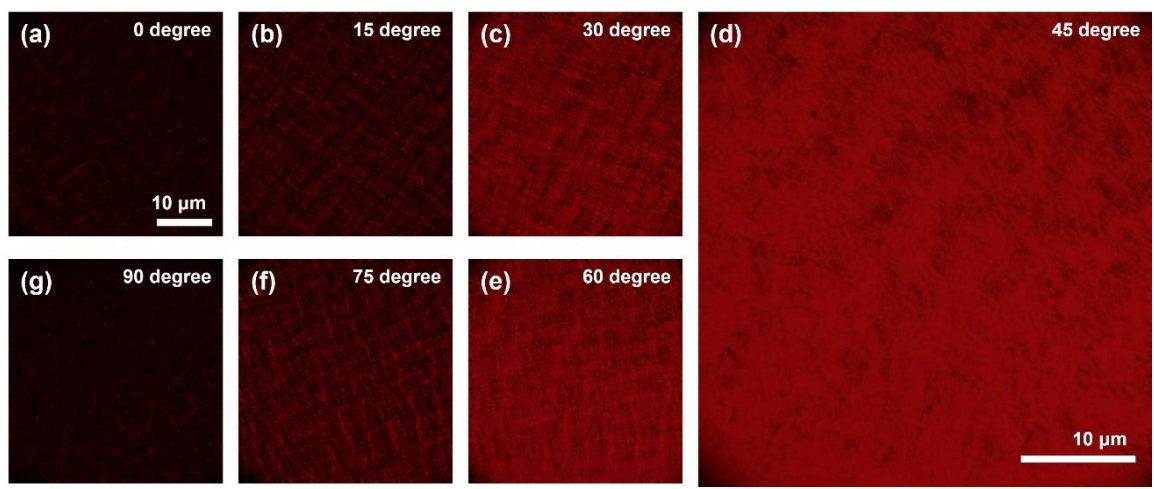

Supplementary Fig. 6 POM photographs of the 90 mg ml$^{-1}$ BBL$_{99}$-MSA solutions.

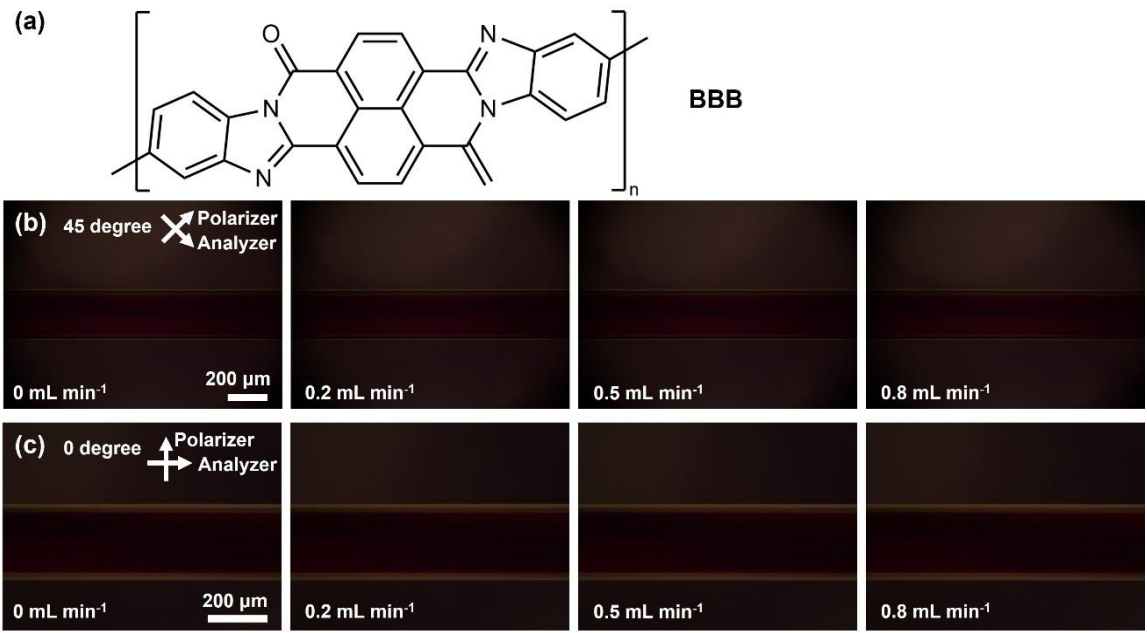

Supplementary Fig. 7 POM photographs of the BBB-MSA.

(a) The chemical structure of BBB. (b-c)POM images of the BBB-MSA solutions in a 200 μm-diameter glass tube in pressure-driven nematic flows at 45°(b) and 0°(c) to the analyzer direction with 0, 0.2, 0.5, 0.8 mL min$^{-1}$.

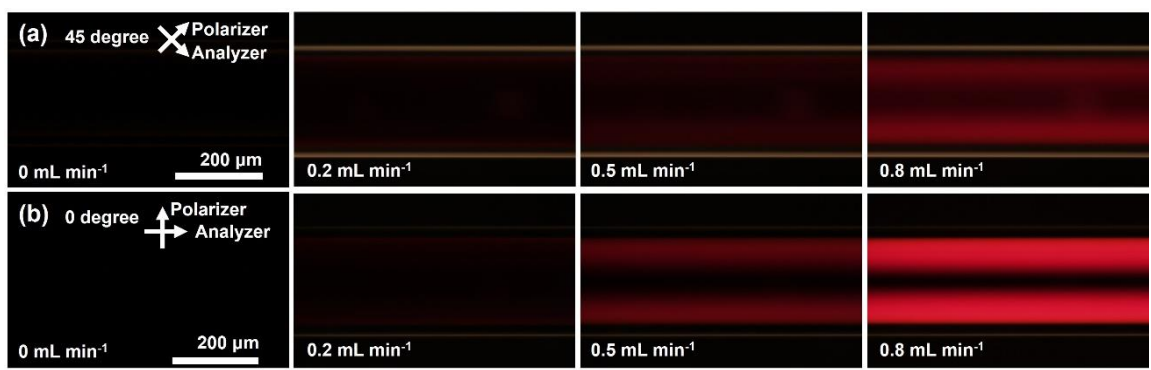

Supplementary Fig. 8 POM images of the BBL$_{39}$-MSA solutions

POM images of the BBL$_{39}$-MSA solutions in a 200 mm-diameter glass tube in pressure-driven nematic flows at 45° (a) and 0° (b) to the analyzer direction with 0, 0.2, 0.5, 0.8 mL min$^{-1}$.

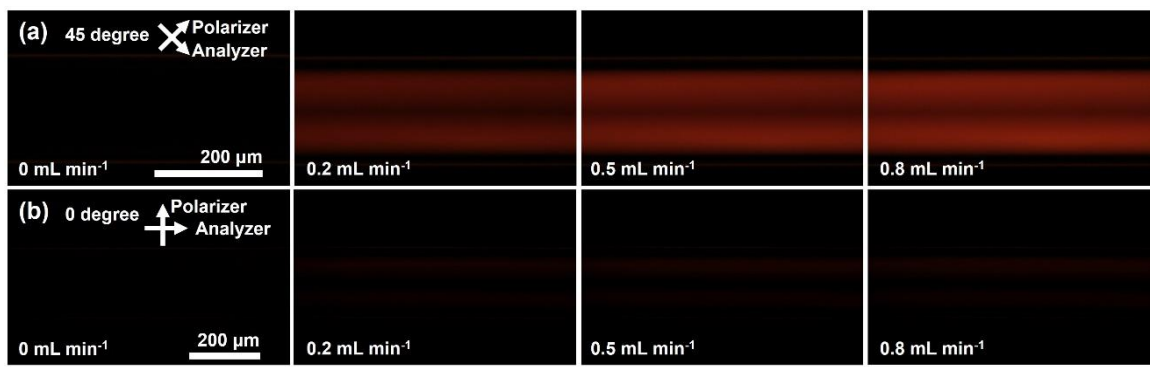

Supplementary Fig. 9 POM images of the PBFDO-DMSO solutions.

POM images of the PBFDO-DMSO solutions in a 200 mm-diameter glass tube in pressure-driven nematic flows at 45° (a) and 0° (b) to the analyzer direction with 0, 0.2, 0.5, 0.8 mL min$^{-1}$.

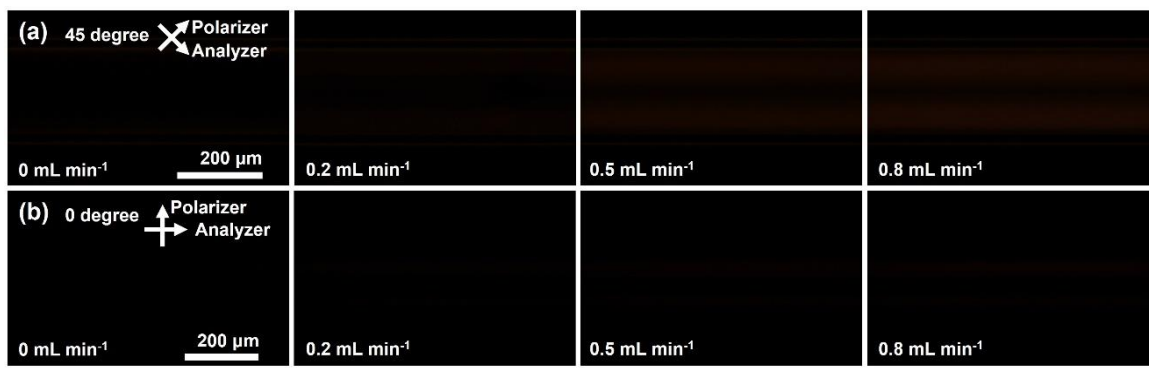

Supplementary Fig. 10 POM images of the PEDOT:PSS-H$_2$O solutions.

POM images of the PEDOT:PSS-H$_2$O solutions in a 200 mm-diameter glass tube in pressure-driven nematic flows at 45° (a) and 0° (b) to the analyzer direction with 0, 0.2, 0.5, 0.8 mL min$^{-1}$.

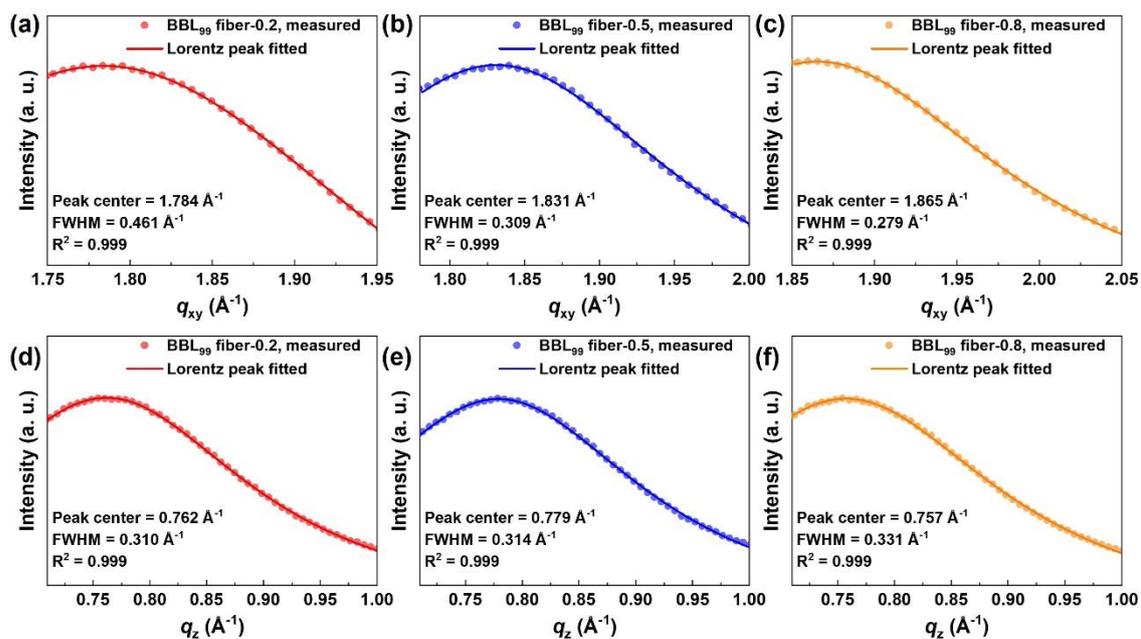

Supplementary Fig. 11 GIWAXS analysis of BBL$_{99}$ fiber.

(a-c) π-π stacking (010) diffraction analysis of BBL$_{99}$ fiber-0.2 (a), BBL$_{99}$ fiber-0.5 (b), and BBL$_{99}$ fiber-0.8 (c). d-e) Lamellar (100) peak analysis of BBL$_{99}$ fiber-0.2 (a), BBL$_{99}$ fiber-0.5 (b), and BBL$_{99}$ fiber-0.8 (c).

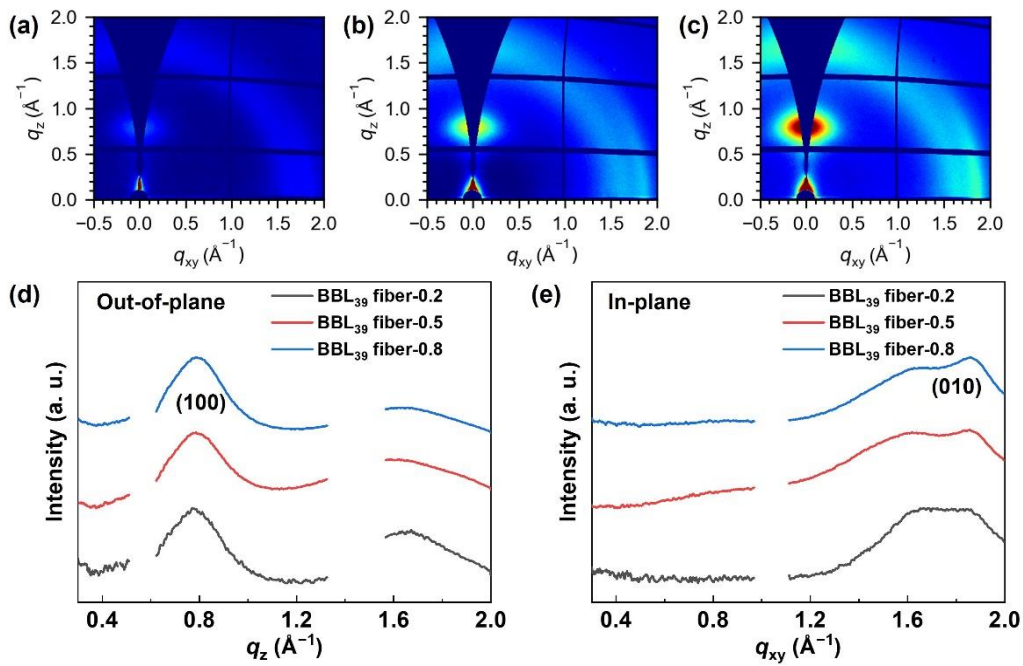

Supplementary Fig. 12 GIWAXS analysis of BBL$_{39}$ fiber.
GIWAXS patterns of BBL$_{39}$ fiber-0.2 (a), BBL$_{39}$ fiber-0.5 (b), and BBL$_{39}$ fiber-0.8 (c). (d,e) In-plane (d) and out-of-plane (e) GIWAXS line cuts of BBL$_{39}$ fiber-0.2, BBL$_{39}$ fiber-0.5, and BBL$_{39}$ fiber-0.8.

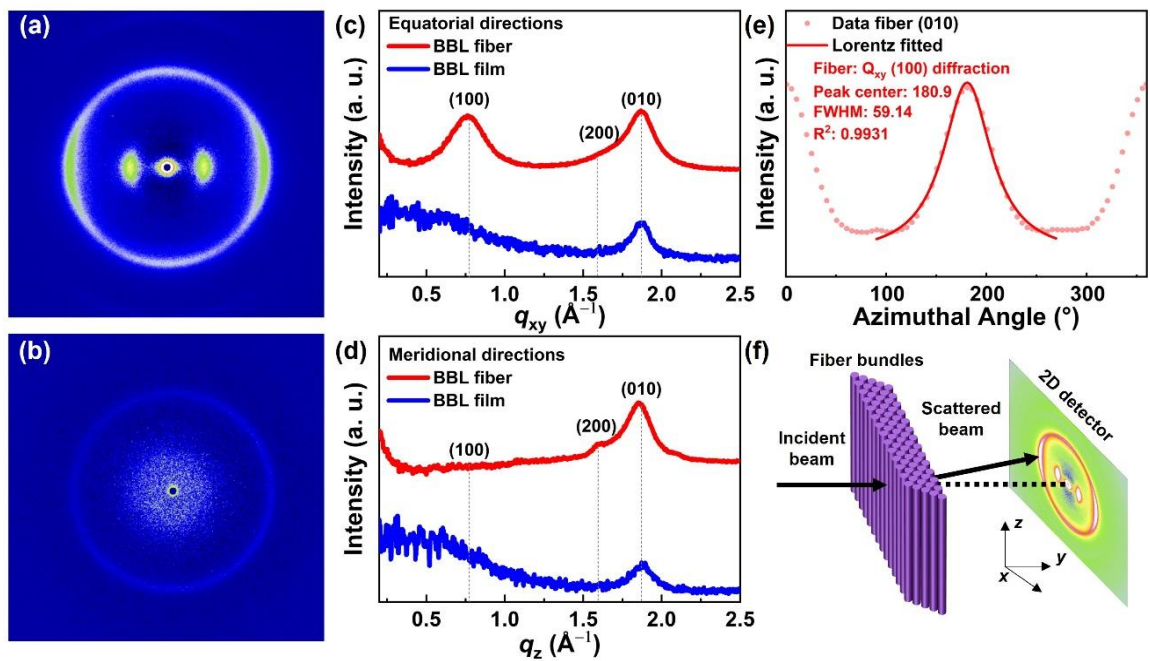

Supplementary Fig. 13 The WAXS of BBL$_{99}$ fiber and film.

(a, b) 2D WAXS patterns of BBL$_{99}$ fiber (a) and BBL$_{99}$ film (b). (c, d) Equatorial (c) and meridional (d) directions WAXS line cuts of BBL fiber and film. (e) Azimuthal profiles of (100) reflections were obtained from the WAXS diagram of BBL$_{99}$ fiber. (f) Schematic diagram of fiber WAXS measurement.

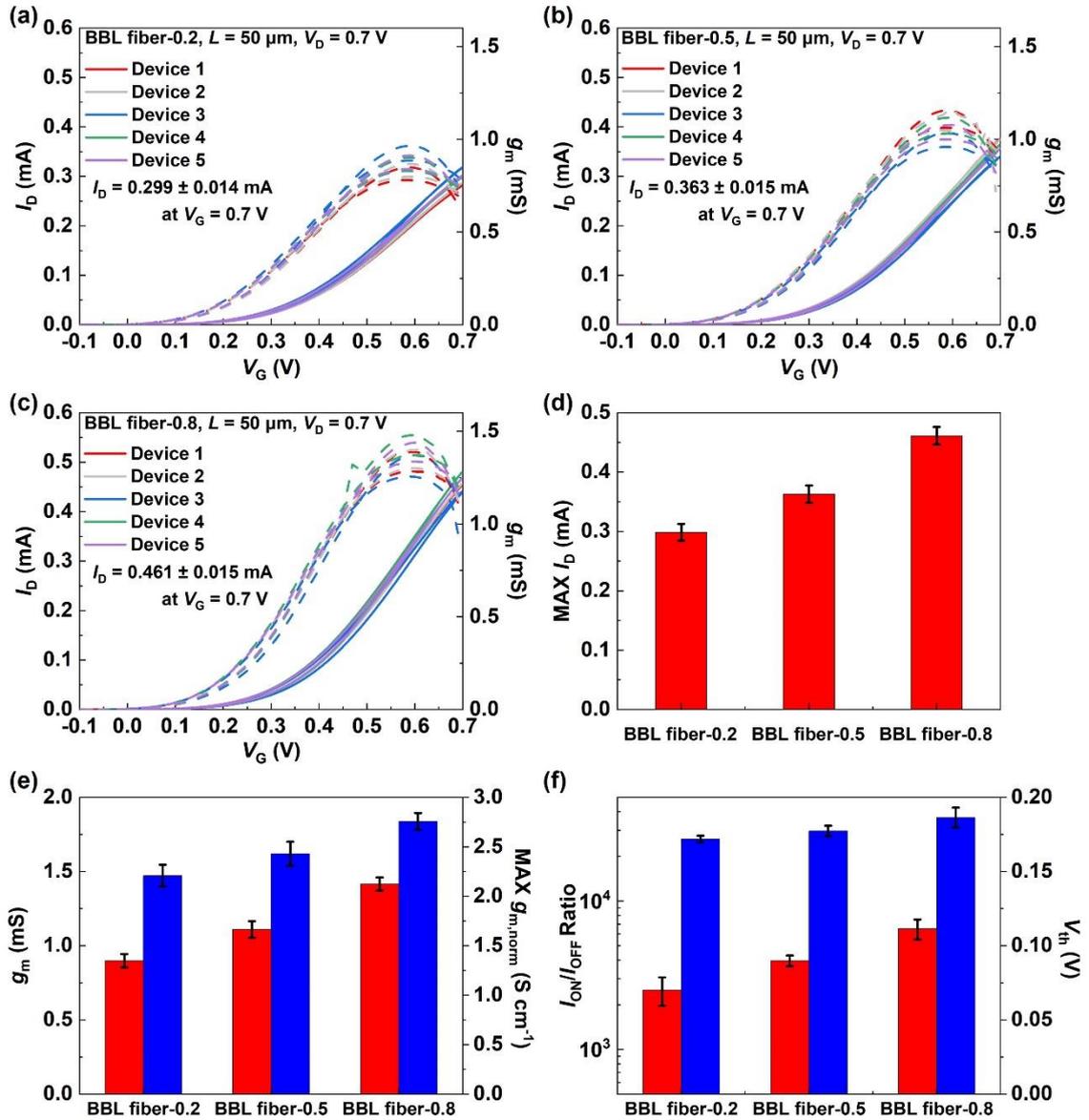

Supplementary Fig. 14 OECT performance of BBL$_{99}$ fiber.

(a-c) Transfer characteristics of five different OECTs based on BBL$_{99}$ fiber-0.2 (a), BBL$_{99}$ fiber-0.5 (b), and BBL$_{99}$ fiber-0.8 (c). All OECTs have the same channel geometry ($L$ = 50 μm, A was defined by the size of BBL$_{99}$ fibers). (d) All BBL$_{99}$-based OECTs show excellent reproducibility, with a standard deviation lower than 3%.

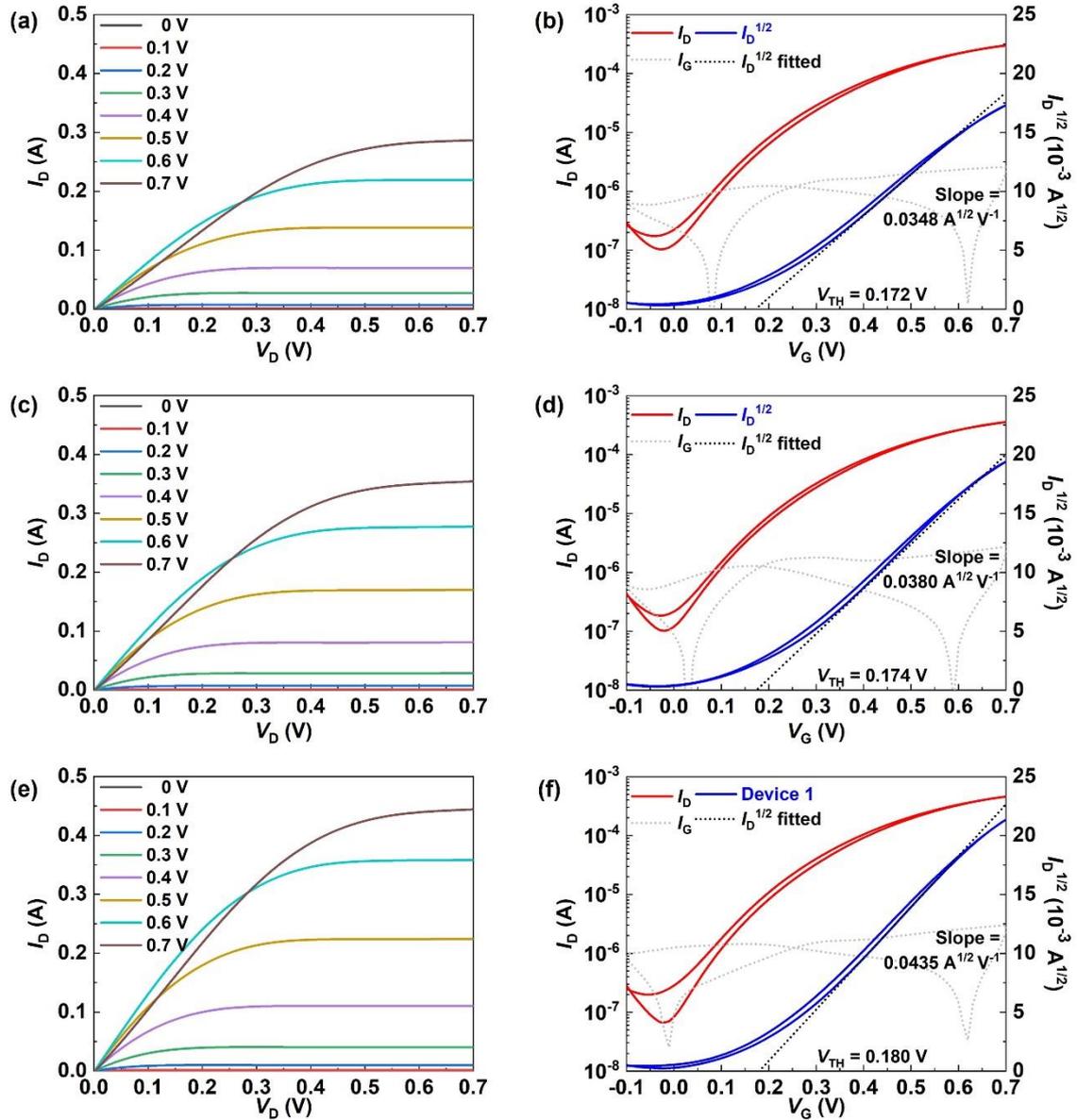

Supplementary Fig. 15 OECT performance of BBL$_{99}$ fiber.

(a-f) Output (a, c, e) and transfer (b, d, f) curves of BBL fiber-0.2 (a, b), BBL fiber-0.5 (c, d), and BBL fiber-0.8 (e, f). Threshold voltages ($V_{th}$) of the four BBLs are also reported. All OECTs have the same channel geometry ($L$ = 50 μm, $A$ was defined by the size of BBL fibers).

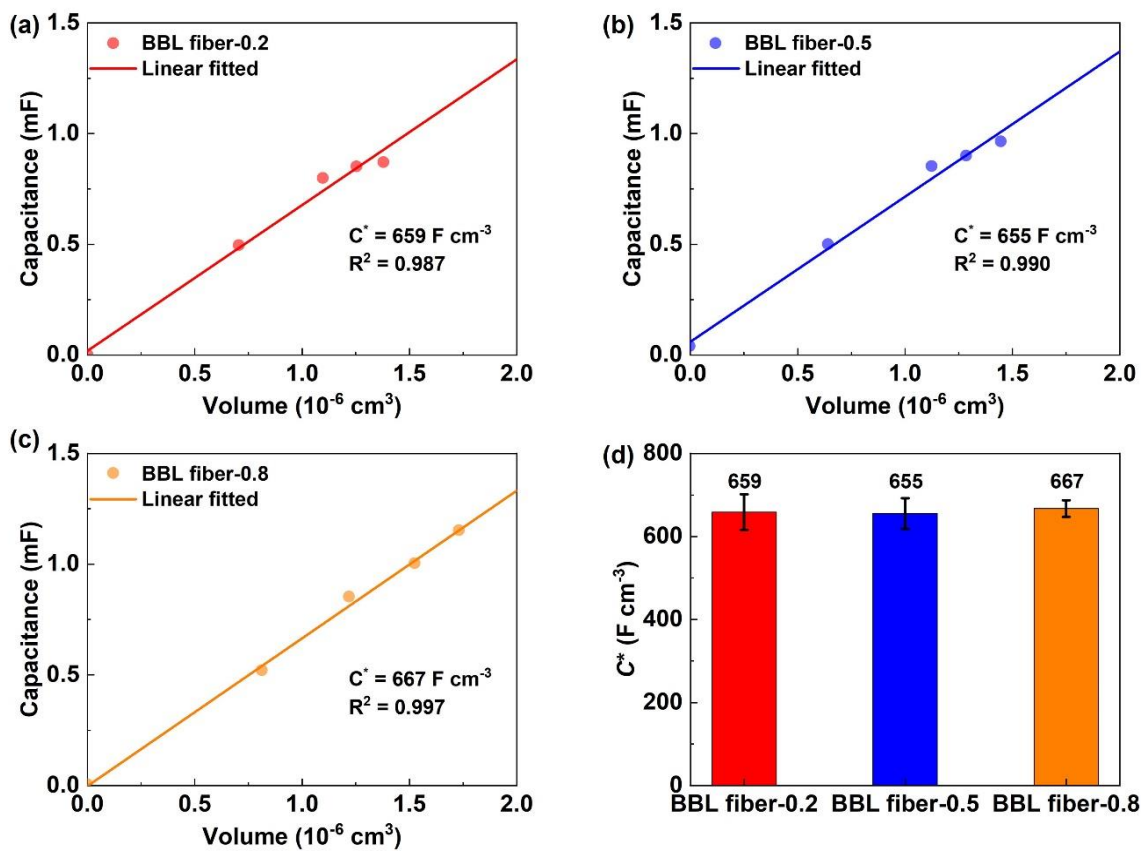

Supplementary Fig. 16 Volumetric capacitance of BBL$_{99}$ fiber.

(a-c) Volume-dependent capacitance of BBL fiber-0.2 (a), BBL fiber-0.5 (b), and BBL fiber-0.8 (c). (d) Summary of volumetric capacitance. The capacitance values were ascertained through analysis of the electrochemical impedance spectroscopy (EIS) spectra, and the volumetric capacitance was determined by employing linear regression on the capacitance data as a function of volume.

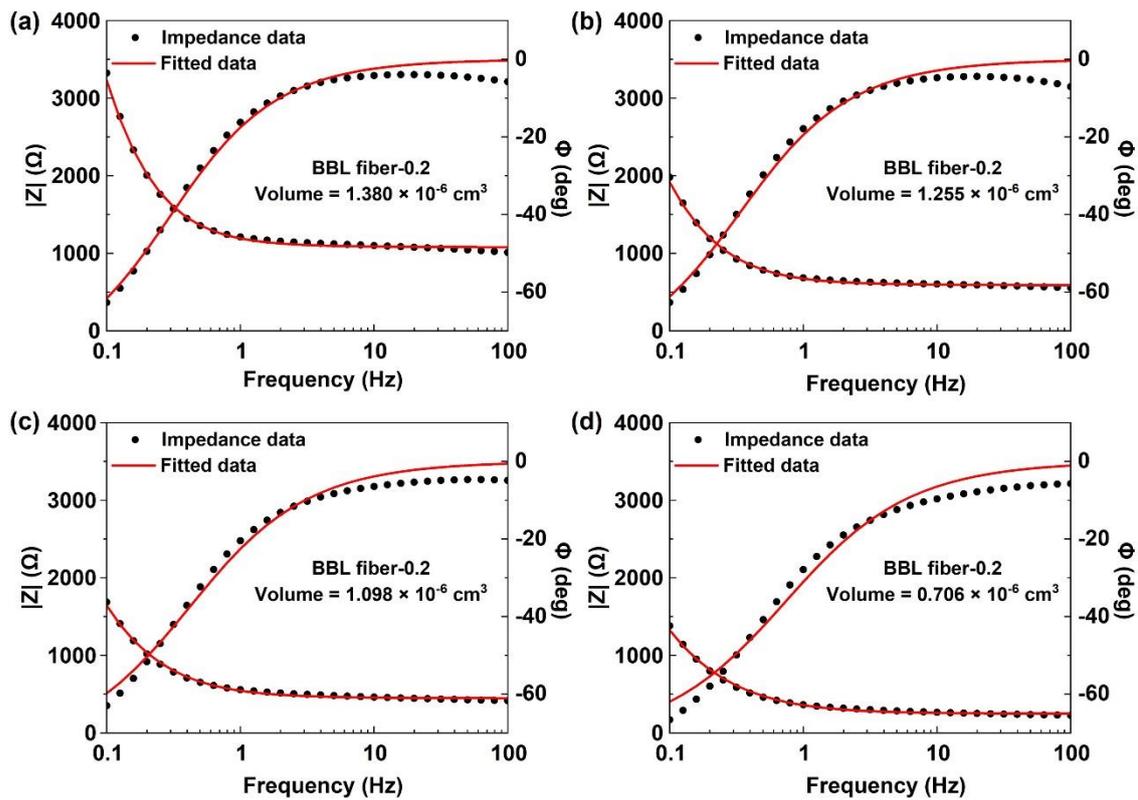

Supplementary Fig. 17 Electrical impedance spectroscopy of $BBL_{99}$ fiber-0.2. Electrical impedance spectroscopy was carried out on different volumes of $BBL_{99}$ fibre-0.2 placed on a gold electrode as a working electrode. The complex impedance measurements obtained were fitted to the *Rs(Rp||Q)* equivalent circuit.

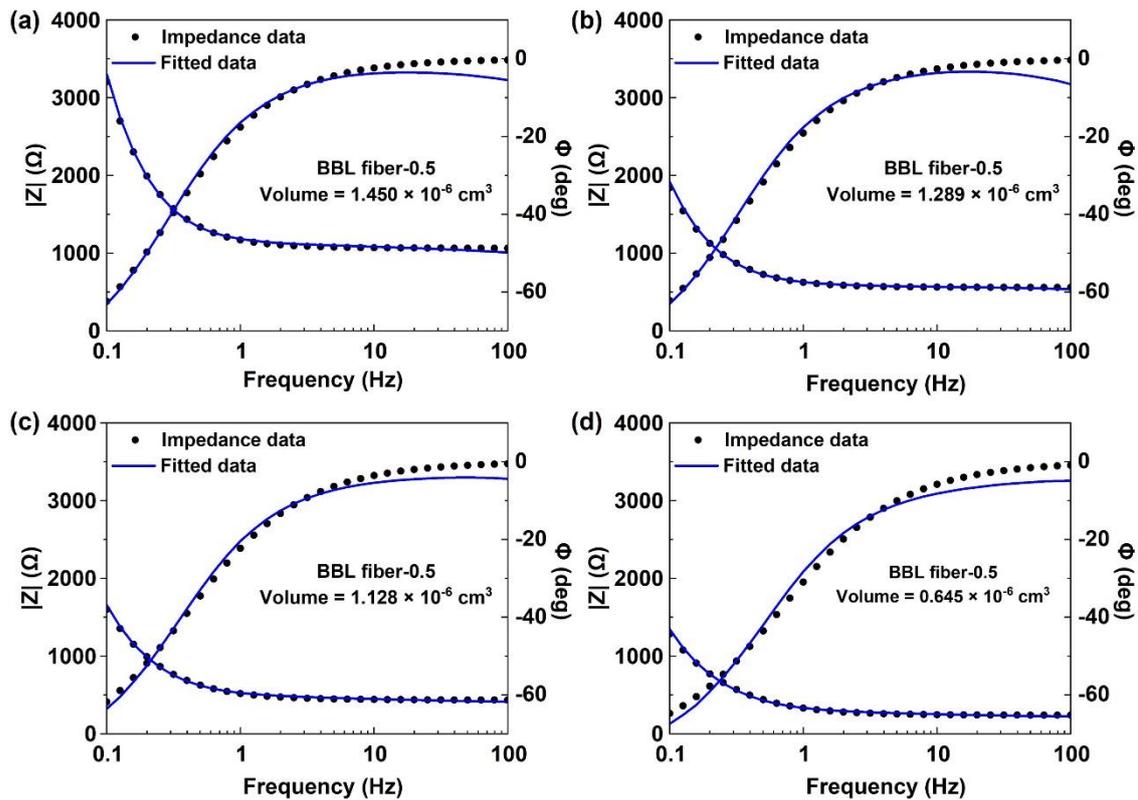

Supplementary Fig. 18 Electrical impedance spectroscopy of $BBL_{99}$ fiber-0.5. Electrical impedance spectroscopy was carried out on different volumes of $BBL_{99}$ fibre-0.5 placed on a gold electrode as a working electrode. The complex impedance measurements obtained were fitted to the *Rs(Rp||Q)* equivalent circuit.

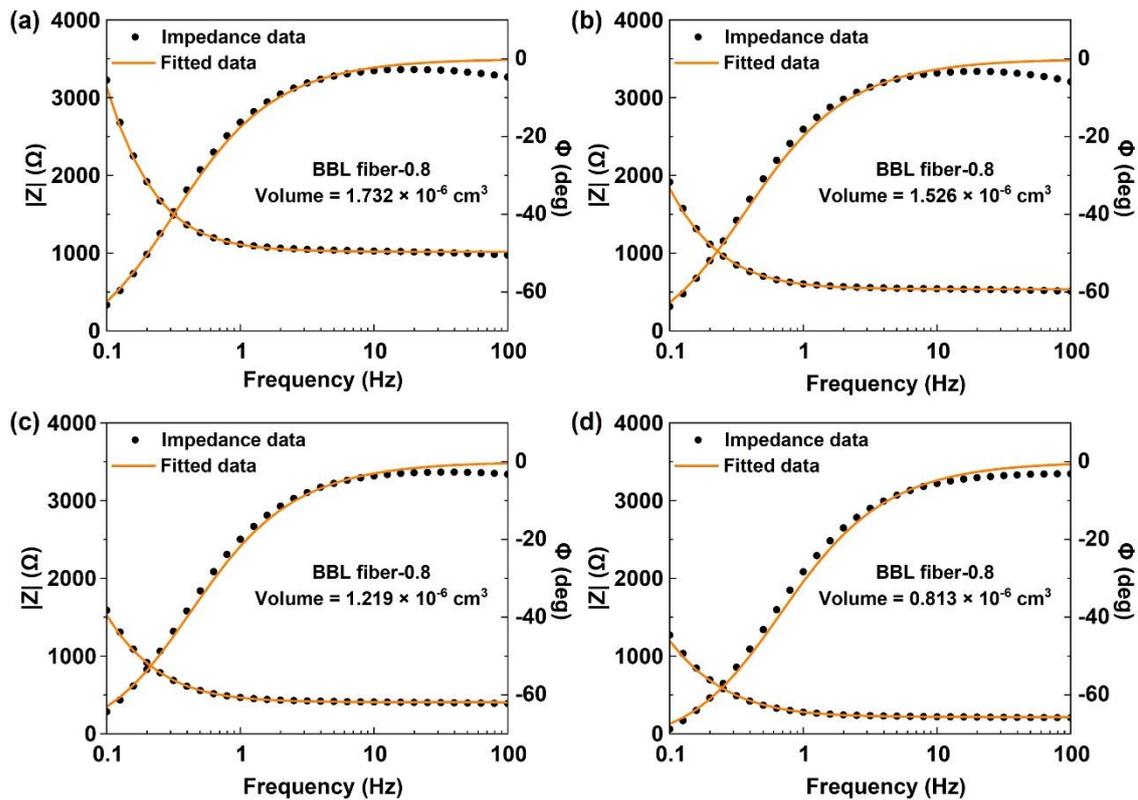

Supplementary Fig. 19 Electrical impedance spectroscopy of BBL$_{99}$ fiber-0.8. Electrical impedance spectroscopy was carried out on different volumes of BBL$_{99}$ fibre-0.8 placed on a gold electrode as a working electrode. The complex impedance measurements obtained were fitted to the *Rs(Rp||Q)* equivalent circuit.

Supplementary Table 4 Summary of BBL$_{99}$ fiber OECT.

| Material | L (μm) | Diameter of fibers (μm) | A (μm$^2$) | $g_{m,norm}$ (S cm$^{-1}$) | $V_{th}$ (V) | $I_{ON}/I_{OFF}$ (× 10$^3$) |
|---|---|---|---|---|---|---|
| BBL$_{99}$ fiber-0.2 | 50 | 16.096 ± 0.093 | 203.482 | 2.209 ± 0.110 | 0.172 ± 0.002 | 2.517 ± 0.544 |
| BBL$_{99}$ fiber-0.5 | 50 | 17.056 ± 0.128 | 228.478 | 2.429 ± 0.122 | 0.177 ± 0.004 | 3.962 ± 0.330 |
| BBL$_{99}$ fiber-0.8 | 50 | 18.084 ± 0.113 | 256.850 | 2.756 ± 0.086 | 0.186 ± 0.007 | 6.508 ± 0.100 |

Supplementary Table 5 Summary of BBL$_{99}$ fiber OECT performance

| Material | $L$ (μm) | Diameter of fibers (μm) | $\mu C^*$ (F cm$^{-1}$ V$^{-1}$ s$^{-1}$) | $C^*$ (F cm$^{-3}$) | $\mu$ (cm$^2$ V$^{-1}$ s$^{-1}$) |
|---|---|---|---|---|---|
| BBL$_{99}$ fiber-0.2 | 50 | 16.096 ± 0.093 | 5.911 ± 0.179 | 659 ± 43 | (8.975 ± 0.646) × 10$^{-3}$ |
| BBL$_{99}$ fiber-0.5 | 50 | 17.056 ± 0.128 | 6.594 ± 0.316 | 655 ± 27 | (1.006 ± 0.075) × 10$^{-2}$ |
| BBL$_{99}$ fiber-0.8 | 50 | 18.084 ± 0.113 | 7.659 ± 0.484 | 667 ± 20 | (1.153 ± 0.080) × 10$^{-2}$ |

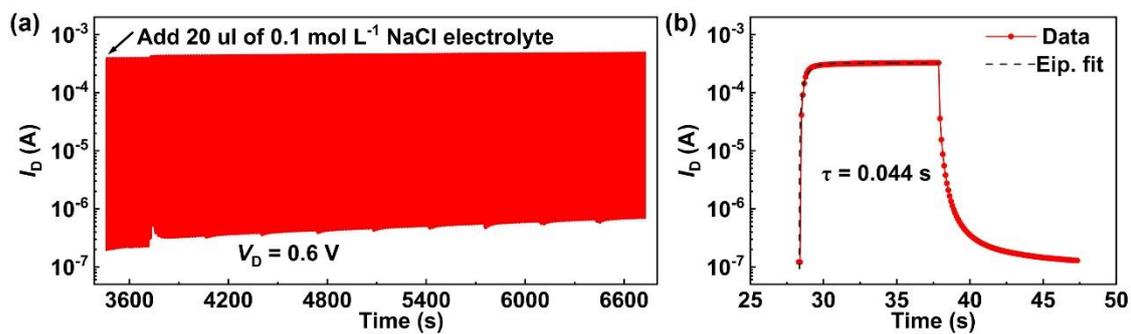

Supplementary Fig. 20 Stability and temporal response of BBL$_{99}$ fiber OECT.

(a) After nearly one hour of stability test (**Fig. 3d**), 20 μL of 0.1 mol L$^{-1}$ NaCl electrolyte was added again for the stability test. The BBL fiber OECT still maintained excellent cycle stability; (b) Temporal response of the drain current as reported in the main text. The dashed black line was an exponential fit with 0.044 s for the turn-on.

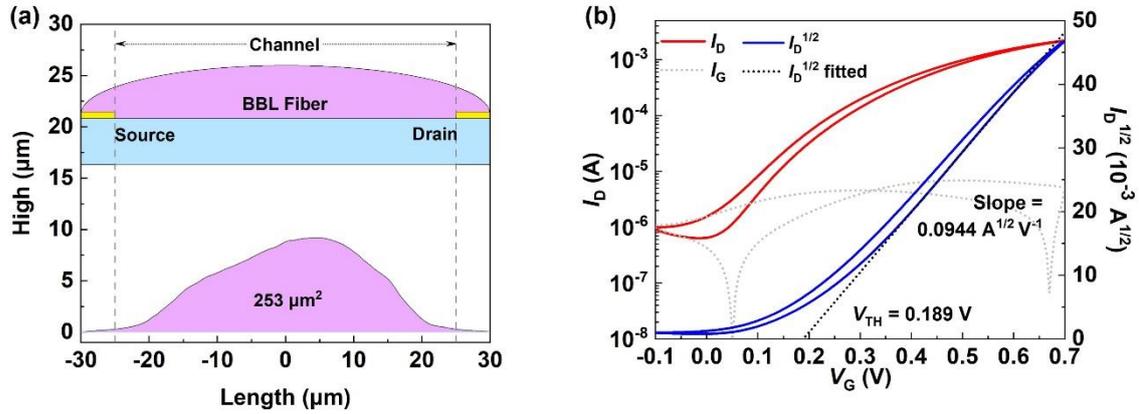

Supplementary Fig. 21 BBL$_{99}$ fiber radial OECTs.

(a) Schematic of the radial OECT; (b) Cross-sectional area of the BBL fiber-0.8. Top: cross-sectional schematic of the radial OECT; (c) Transfer characteristics of the radial OECT and the corresponding transconductance values at $V_d$ = 0.7 V; (d) Transfer curves of radial BBL$_{99}$ fiber-0.8; (e) $\mu C^*$ and $\mu_{OECT}$ values in axial and radial directions for the BBL$_{99}$ fiber-0.8; (f) $g_{m,\ norm}$ and $V_{th}$ values in axial and radial directions for the BBL$_{99}$ fiber-0.8;

Supplementary Table 6 The electrical properties in fiber-based OECTs

| Ref No. | Channel | On/Off ratio | Drive (V) | $g_m$ (µS) |
|---|---|---|---|---|
| Ref. S[1] | PEDOT/PSS | $10^3$ | 1 | 1000 |
| Ref. S[2] | PPy/PVA/PE | $2.6 \times 10^2$ | 3 | - |
| Ref. S[3] | PPy | $10^4$ | 2 | - |
| Ref. S[4] | CNT | $10^2$ | 1 | 1350 |
| Ref. S[5] | PPy/Graphene | $10^2$ | 2 | - |
| Ref. S[6] | PAni | $10^3$ | 0.6 | 60 |
| This work | BBL | $10^3$ | 0.6 | 1101 |

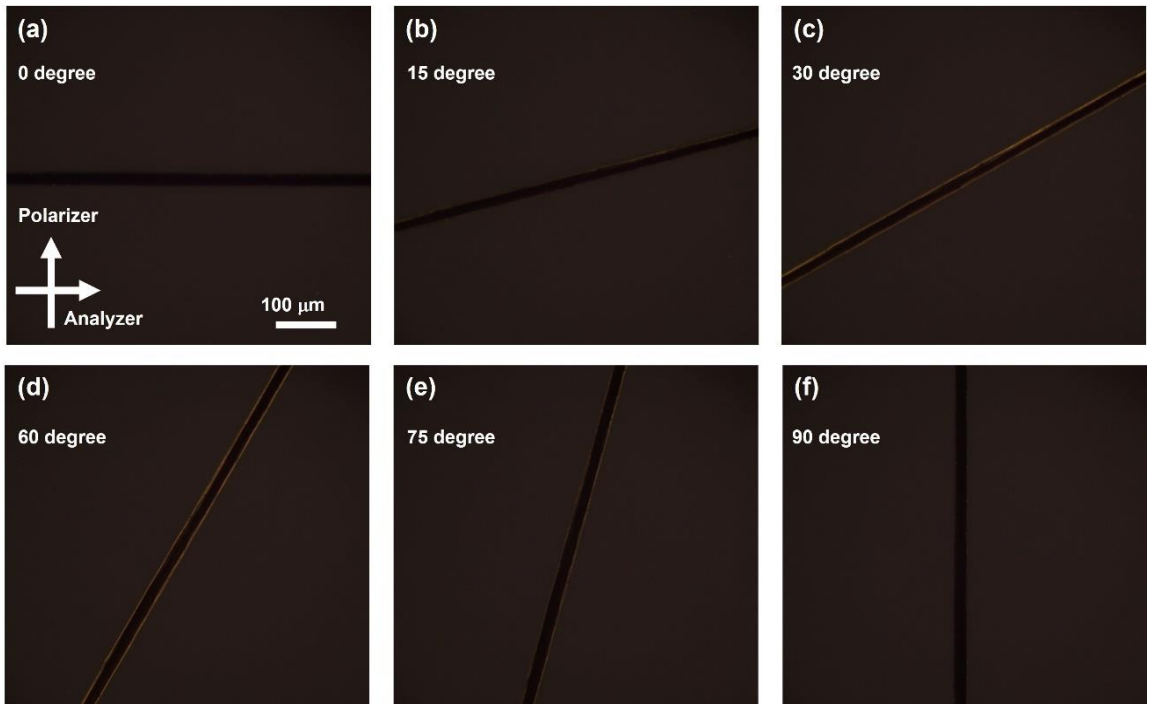

Supplementary Fig. 22 The POM image of the BBL$_{99}$ fiber.

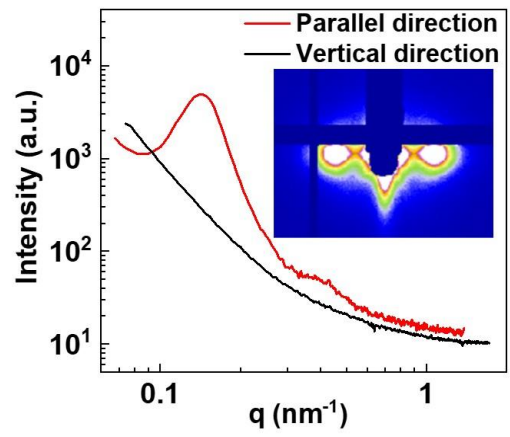

Supplementary Fig. 23 SAXS of BBL$_{99}$ fiber-0.8.

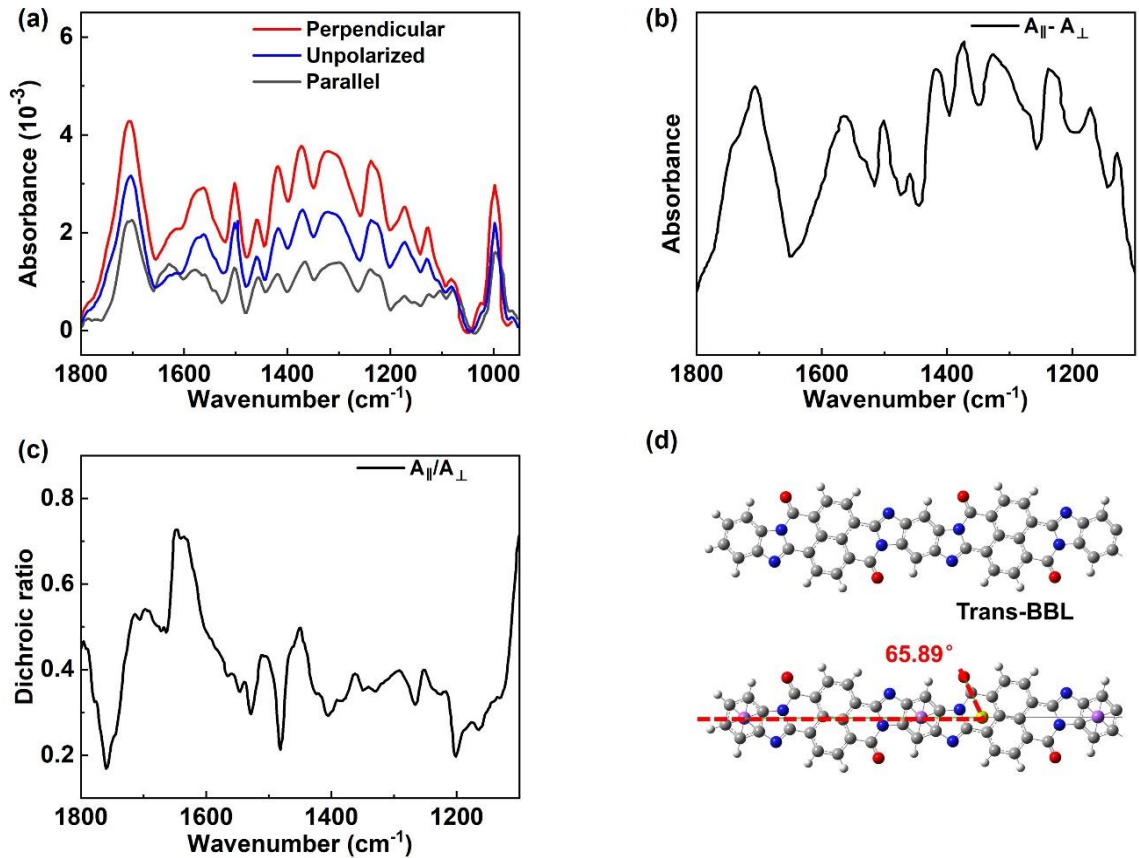

Supplementary Fig. 24 Polarized FTIR of BBL$_{99}$ fibers.

(a) Polarized FTIR of BBL fibers; The dichroic difference (a) and the dichroic ratio (b) of BBL fiber; (c) The direction schematic diagram of *trans*-BBL and the direction of BBL molecular chain and C=O bond.

The dichroic difference is calculated as $R = A_{\parallel} - A_{\perp}$; The dichroic ratio is calculated as $R = A_{\parallel}/A_{\perp} = 0.52$; The factor of orientation is calculated as $f = \frac{(R-1)(R_0+2)}{(R+2)(R_0-1)}$ ; Where $A$ is the absorbance of the fibers with parallel ($\parallel$) or perpendicular ($\perp$) light; The vibration of the C=O bond (1705 cm$^{-1}$) is at a roughly 65.89-degree angle ($\alpha$) from the chain propagation direction. The dichroic ratio of a perfectly oriented polymer is calculated as $R_0 = 2\cot^2\alpha = 0.40$; The factor of orientation is 0.76.

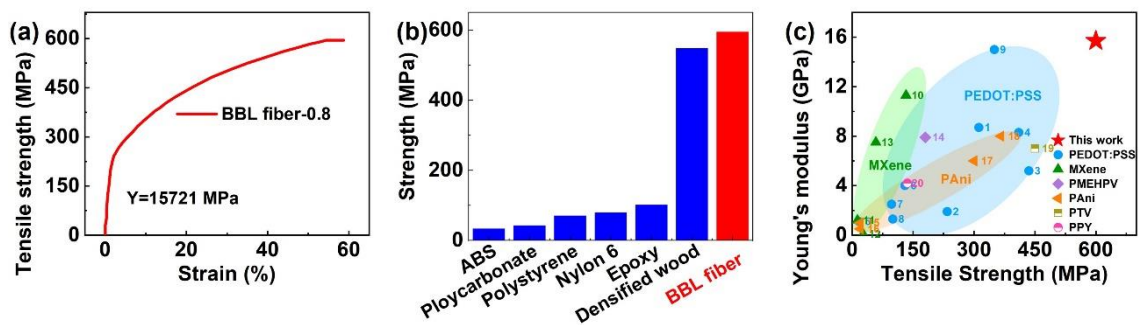

Supplementary Fig. 25 Mechanical properties of semiconductor fibers. (a) Stress-strain curve of BBl$_{165}$ fiber with a diameter of 20 μm showed that the tensile strength was 595.1 MPa and Young's modulus (Y) was 15.7 GPa; (b) Comparison of the tensile strength of BBL fibers (595.1 MPa) with other broadly applied polymer-based materials and Hu's densified wood. (c) Compares Young's modulus and tensile strength of manufactured BBL fibers with MXene composite fibers, PEDOT: PSS fibers, and other conjugate polymer fibers fabricated in previous studies[1,7–25]. The Ashby plot indicated that wet-spun pure BBL fibers outperformed the other mentioned fibers regarding Young's modulus and tensile strength, implying that the BBL fibers had a well-mechanical property.

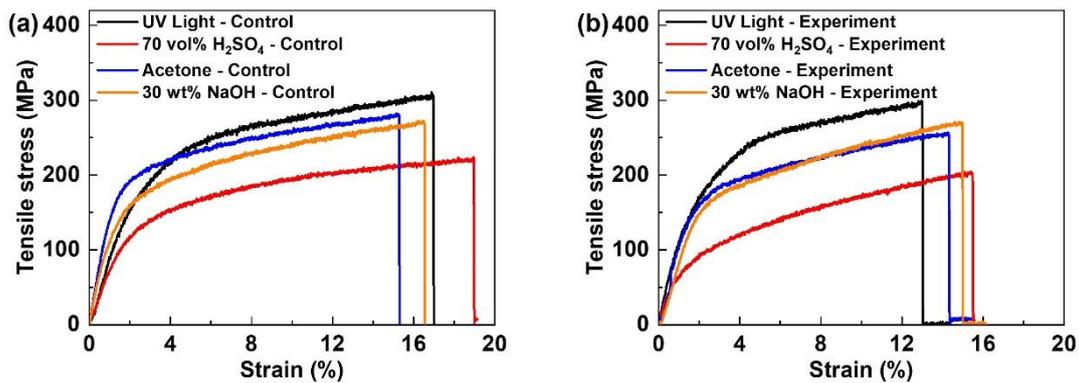

Supplementary Fig. 26 Environmental stability of BBL$_{99}$ fibers. Stress-strain curve of BBL$_{99}$ fibers before and after ultraviolet radiation (365 nm LED, 5000 W m$^{-2}$, 10 cm distance) for 12 h (a-b), soaked in acetone solution(AR) for 24 h (c-d), soaked in 30 wt% NaOH solution for 24 h (e-f) and soaked in 90 vol% H$_2$SO$_4$ solution for 24 h (g-h).

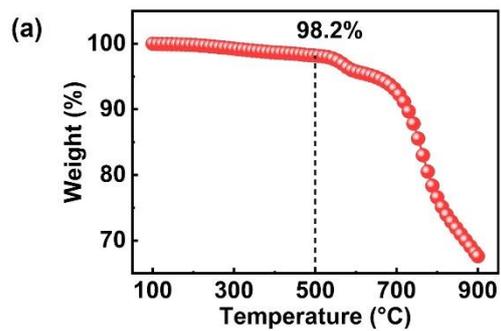
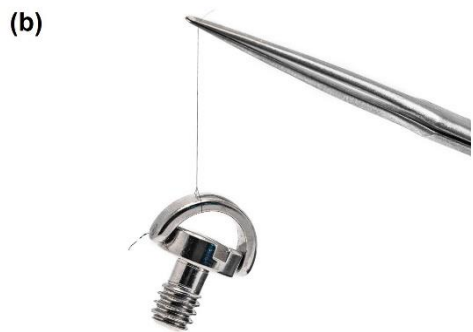
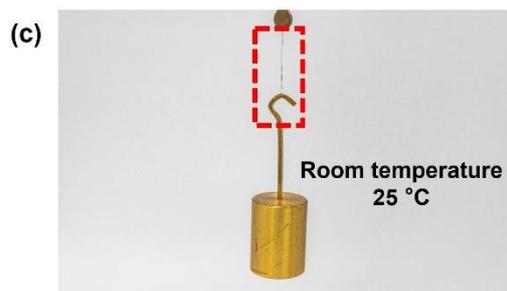
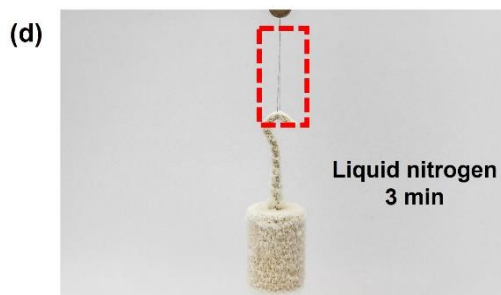

Supplementary Fig. 27 Thermomechanical properties of BBL$_{99}$ fibers.

(a)TGA curves of BBL (20 μm) at 10 K min$^{-1}$ under nitrogen atmosphere. Digital photo of BBL fibers (20 μm) hanging with a bolt (b) and a 10 g weight at room temperature (c) and in liquid nitrogen for 3 minutes (d).

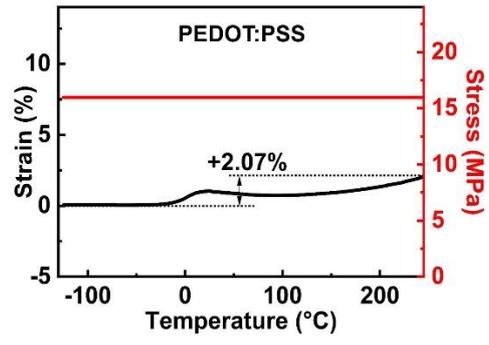

Supplementary Fig. 28 DMA graph of PEDOT:PSS fiber.

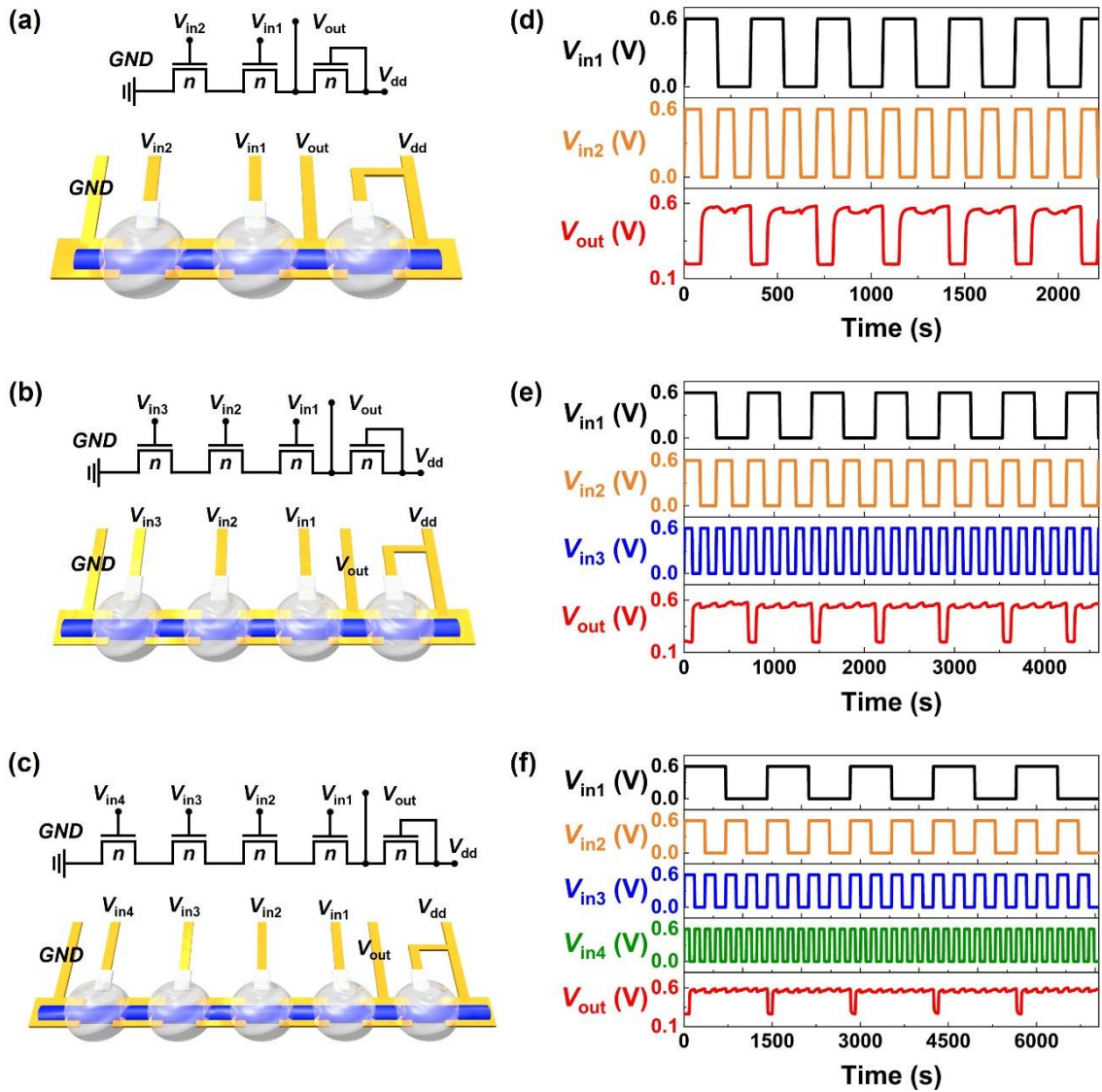

Supplementary Fig. 29 Single semiconductor fiber NAND logic circuits.

A schematic diagram and logic circuit design of the single fiber NAND of 2 (a), 3 (b), and 4 (c) input signals. Output characteristics of the single fiber NAND of 2 (d), 3 (e), and 4 (f) input signals.